 \documentclass[preprint2]{aastex}

\shorttitle{Kinematics of \ion{C}{2} $\lambda$ 6578}
\shortauthors{Richer et al.}

\begin{document}


\title{The Kinematics of the Permitted \ion{C}{2} $\lambda$ 6578 Line in a Large Sample of Planetary Nebulae\footnote{Based upon observations carried out at the Observatorio Astron\'omico Nacional on the Sierra San Pedro M\'artir (OAN-SPM), Baja California, M\'exico.}\ $^{\mathrm ,\,}$\footnote{Based on observations made with the NASA/ESA Hubble Space Telescope, and obtained from the Hubble Legacy Archive, which is a collaboration between the Space Telescope Science Institute (STScI/NASA), the Space Telescope European Coordinating Facility (ST-ECF/ESA) and the Canadian Astronomy Data Centre (CADC/NRC/CSA).}}


\author{Michael G. Richer, Genaro Su\'arez, Jos\'e Alberto L\'opez, and Mar\'\i a Teresa Garc\'\i a D\'\i az}
\affil{Instituto de Astronom\'\i a, Universidad Nacional Aut\'onoma de M\'exico, Ensenada, Baja California, Mexico}
\email{richer@astrosen.unam.mx, gsuarez@astro.unam.mx, jal@astrosen.unam.mx, tere@astro.unam.mx}







\begin{abstract}
We present spectroscopic observations of the \ion{C}{2} $\lambda$6578 permitted line for 83 lines of sight in 76 planetary nebulae at high spectral resolution, most of them obtained with the Manchester Echelle Spectrograph on the 2.1\,m telescope at the Observatorio Astron\'omico Nacional on the Sierra San Pedro M\'artir.  We study the kinematics of the \ion{C}{2} $\lambda$6578 permitted line with respect to other permitted and collisionally-excited lines.  Statistically, we find that the kinematics of the \ion{C}{2} $\lambda$6578 line are not those expected if this line arises from the recombination of C$^{2+}$ ions or the fluorescence of C$^+$ ions in ionization equilibrium in a chemically-homogeneous nebular plasma, but instead its kinematics are those appropriate for a volume more internal than expected.  The planetary nebulae in this sample have well-defined morphology and are restricted to a limited range in H$\alpha$ line widths (no large values) compared to their counterparts in the Milky Way bulge, both of which could be interpreted as the result of young nebular shells, an inference that is also supported by nebular modeling.  Concerning the long-standing discrepancy between chemical abundances inferred from permitted and collisionally-excited emission lines in photoionized nebulae, our results imply that multiple plasma components occur commonly in planetary nebulae.  
\end{abstract}





\section{Introduction}
\label{sec_introduction}

The chemical composition of matter throughout the universe is routinely determined via observations of ionized gas.  This is most easily done using the brightest emission lines, which are collisionally-excited (forbidden) emission lines of heavy elements and permitted lines of hydrogen and helium.  However, it is also feasible, at least in principle in \ion{H}{2} regions and planetary nebulae, to use the much fainter permitted lines of the heavy elements.  When abundances inferred from both types of heavy element lines are compared, the abundances derived from the permitted lines are almost invariably larger.  The difference between the abundances calculated from both types of lines is usually about a factor of two in \ion{H}{2} regions and in about 80\% of planetary nebulae, but exceeds a factor of five in approximately 20\% of planetary nebulae with record values in excess of 100 \citep[e.g.,][]{garciarojasesteban2007,liu2010,corradietal2015}.  This difference in abundances is known as the \lq\lq abundance discrepancy" in ionized nebulae and is usually characterized by the abundance discrepancy factor (ADF), which is the ratio of the abundances calculated from permitted lines with respect to the abundances inferred from collisionally-excited lines.  While the abundance discrepancy is disconcerting, heavy element ratios, such as $\mathrm N/\mathrm O$ or $\mathrm C/\mathrm O$, are not affected provided the same type of line is used for both elements \citep[e.g.,][]{liu2006}.  

Since attention was first directed to this issue \citep{wyse1942}, various mechanisms to explain the abundance discrepancy have been proposed.  Those that have been best developed focus on the temperature distribution and the appropriate temperatures to use \citep{peimbert1967, estebanetal2004, nichollsetal2012} or the existence of multiple plasma components \citep{liuetal2000}.  However, other ideas have been considered, such as recombination efficiency \citep{garnettdinerstein2001, rodriguezgarciarojas2010} or density structures \citep{viegasclegg1998,tsamisetal2011, mesadelgadoetal2012}, while other promising possibilities like dust discs \citep{bilikovaetal2012} have yet to be developed.  See \citet{ferland2003}, \citet{peimbertpeimbert2006}, \citet{liu2006, liu2010}, \citet{bohigas2009}, or \citet{ferlandetal2016} for more complete discussions.  

Recently, \citet{richeretal2013} presented a detailed study of the kinematics of many types of emission lines in the planetary nebula NGC 7009, whose ADF is a factor of five \citep{liuetal1995, luoetal2001, fangliu2013}.  \citet{richeretal2013} find that the kinematics of the optical permitted lines in NGC 7009 indicate the presence of a plasma component emitting in these lines whose position within the nebula does not agree with that expected given the observed ionization structure of this object.  This anomalous component of the optical permitted lines arises from a volume of the nebula internal to that expected from ionization equilibrium in a chemically-homogeneous plasma.  This result coincides with several studies based upon the spatial distribution of the emission from permitted and collisionally-excited emission lines \citep{barker1982, barker1991, liuetal2000, garnettdinerstein2001, luoliu2003, tsamisetal2008, corradietal2015, jonesetal2016, garciarojasetal2016a, garciarojasetal2016b}.  These studies find that the emission from permitted lines is more centrally-concentrated than the emission from collisionally-excited lines from the same ions.  

Here, we propose to undertake a more restricted study than \citet{richeretal2013}, but based upon a large sample of planetary nebulae.  Our purpose is to investigate whether the result found by \citet{richeretal2013} occurs commonly in planetary nebulae, since it has a bearing upon which types of solutions to the abundance discrepancy problem are viable generally.  Here, we study the kinematics of a single permitted line from a single heavy element, the \ion{C}{2} $\lambda$6578 line, and compare the results with the kinematics of the H$\alpha$, \ion{He}{2} $\lambda$6560, and [\ion{N}{2}] $\lambda\lambda$6548,6583 lines.  Although it is well-known that the \ion{C}{2} $\lambda$6578 line may be excited indirectly via fluorescence \citep[e.g.,][]{grandi1976}, our analysis accounts for this.  Statistically, we find that the kinematics of the plasma that gives rise to the \ion{C}{2} $\lambda$6578 line is not that expected given the ionization structure of the objects under study.  Hence, we confirm previous results, but with a much larger sample of objects, indicating that multiple plasma components probably occur commonly in planetary nebulae.

\section{Observations, Data Reduction, and Methodology}
\label{section_obs}

\subsection{Observations and Data Reduction}


All of our observations are drawn from the San Pedro M\'artir Kinematic Catalogue of Planetary Nebulae \citep[henceforth, SPM Catalogue;][]{lopezetal2012}\footnote{http://kincatpn.astrosen.unam.mx/}.  Most of these spectra were acquired with the Manchester Echelle Spectrograph \citep[MES:][]{meaburnetal1984, meaburnetal2003} attached to the 2.1\,m telescope at the Observatorio Astron\'omico Nacional on the Sierra San Pedro M\'artir (OAN-SPM).  The MES uses interference filters to isolate the spectral order of interest, order 87 in this case, covering approximately the 6545--6595\AA\ wavelength range and containing the [\ion{N}{2}] $\lambda\lambda$6548,6584, \ion{He}{2} $\lambda$6560 (when present), H$\alpha$, and \ion{C}{2} $\lambda$6578 emission lines.  For most of the observations, a 150\,$\mu$m slit was used, resulting in a spectral resolution of approximately 11\,km\,s$^{-1}$ when converted to velocity.  A few observations were made with narrower slits and had a higher spectral resolution.  Typically, the spatial resolution was 0\farcs6\,pixel$^{-1}$, but, again, was somewhat better for a minority of the observations.  Exposure times were usually of 30 minutes duration.  Spectra of a thorium-argon arc lamp were taken immediately after each object spectrum and allowed wavelength calibration with an internal precision better than 1\,km\,s$^{-1}$.  

\begin{figure}[t]
\includegraphics[width=\columnwidth]{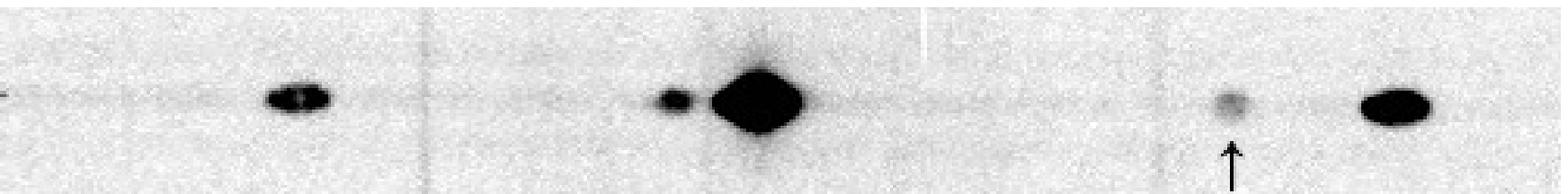}
\includegraphics[width=\columnwidth]{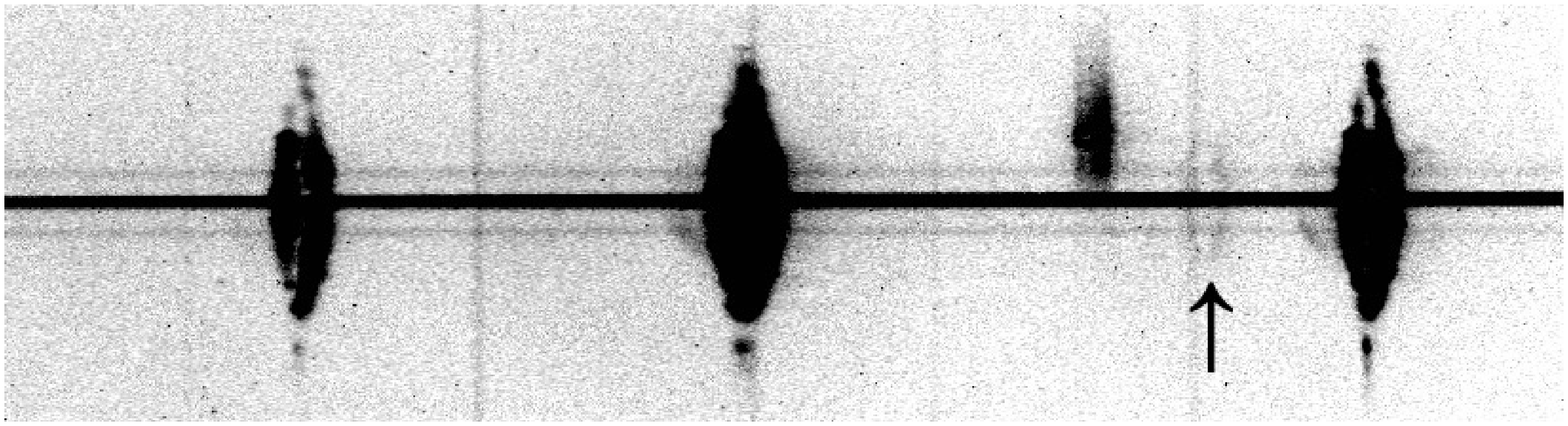}
\caption{These two spectra (M 3-32 top; NGC 40 bottom) illustrate the differences between objects without and with spatial resolution (top and bottom, respectively).  The horizontal axis is the spectral direction and the vertical axis is the spatial direction.  From left to right, the emission lines are [\ion{N}{2}] $\lambda$6548, \ion{He}{2} $\lambda$6560 (only in the top spectrum), H$\alpha$, \ion{C}{2} $\lambda$6578 (arrowed), and [\ion{N}{2}] $\lambda$6583.  Note the ghost between H$\alpha$ and \ion{C}{2} $\lambda$6578 in the bottom spectrum.}
\label{fig_examples}
\end{figure}

Table \ref{table_velocities} presents our sample of objects.  Our sample includes all of those objects in which the \ion{C}{2} $\lambda$6578 emission line was detected.  Since the selection criteria for the SPM Catalogue do not consider the presence of the \ion{C}{2} $\lambda$6578 line or chemical abundances, our sample of objects is blind as respects these selection criteria.  Our sample contains planetary nebulae from both the bulge and disc of the Milky Way and one extragalactic planetary nebula (He 2-436 in the Sagittarius dwarf spheroidal).  There are 83 spectra of 76 individual objects in our sample.  In general, the spectrograph slit was centered upon each object.  For the 7 planetary nebulae with two slit positions, both positions go through the center for IC 4593 and NGC 6543 (different position angles), both are placed symmetrically on either side of the center for NGC 1535 and NGC 7662, while for IC 2149, M 1-42, and NGC 3242 one slit was near the centre and the other slightly farther out (see Table \ref{table_velocities} and the online SPM Catalogue$^3$ for the exact slit positions).  In addition to the objects in Table \ref{table_velocities}, the SPM Catalogue also includes \ion{C}{2} $\lambda$6578 detections for Hu 2-1 and NGC 6891.  We do not include them because the H$\alpha$ line is saturated and so the kinematic and spatial parameters for H$\alpha$ in these objects cannot be estimated properly, and we would have to exclude them from many of the analyses that follow.  

The data were reduced following the standard prescription described by \citet{masseyetal1992} for long slit data, including corrections for image distortion.  When necessary, cosmic rays were removed on a case-by-case basis.  See \citet{lopezetal2012} for a more detailed description of the data reduction.  Figure \ref{fig_examples} presents two representative spectra.  For reference, the line wavelengths that we adopt are 6562.791 \AA\ for H$\alpha$ \citep{cleggetal1999}, 6548.05 \AA\ and 6583.39 \AA\ for [\ion{N}{2}] $\lambda\lambda$6548,6583 \citep{bowen1960}, 6560.1 \AA\ for \ion{He}{2} $\lambda$6560 (Atomic Line List v. 2.05b19/P. van Hoof\footnote{http://www.pa.uky.edu/\%7Epeter/newpage/}), and 6578.05 \AA\ for \ion{C}{2} $\lambda$6578 \citep[][]{ralchenkoetal2011}.




\subsection{Line Widths}\label{sec_methodology}

We characterize the kinematics of the plasma in our sample of planetary nebulae using the widths of the emission lines.  The line widths were measured from one-dimensional spectra extracted from the reduced two-dimensional spectra.  If the object was not resolved spatially, we extracted the entire spatial extent of the object (e.g., Figure \ref{fig_examples}, top panel).  When the object was resolved spatially, we extracted the spatial extent over which the \ion{C}{2} $\lambda$6578 emission was present or the spatial extent over which this emission was best detected.  

When the emission lines in the one-dimensional spectra were symmetric (69 spectra), we fit the observed line profile with a single Gaussian using IRAF's splot task.  The full width at half of maximum intensity (FWHM) of the Gaussian fit to each line was adopted as the line width for each line.  For the spectra in which the line profiles in the extracted spectra were clearly asymmetric or consisted of two components (14 spectra), we also fit these profiles with two Gaussian components using IRAF's splot task.  In this case, we adopted the difference in wavelength between the two Gaussian components as the line width.  We checked that the parameters of the two Gaussian components were stable by varying the initial positions of the two components.  If the two components were not stable, we used the fit of a single Gaussian to the line profile.  Table \ref{table_velocities} lists the observed line widths, $W_{obs}$, for all lines.

For the line widths based upon a single Gaussian fit, we corrected the observed FWHM to account for effects that broaden the lines.  The instrumental broadening (FWHM) was taken as the wavelength interval corresponding to 2.6 pixels, based upon the spectral resolution given in Table \ref{table_diameters}.  We computed the thermal broadening according to $\sigma_{th} = \sqrt{0.00825T_e/A}$, where $A$ is the mass of the ion in atomic mass units, adopting an electron temperature of $T_e=10^4$\,K.  In \S\ref{sec_c2kin}, we show that the correction for thermal broadening to the \ion{C}{2} $\lambda$6578 line widths has no effect upon our conclusions.  Finally, we corrected H$\alpha$ for fine structure broadening using a value of $\sigma^2_{fs} = 10.233$ km$^2$\,s$^{-2}$ \citep{garciadiazetal2008} to account for the fine structure of this line.  The [\ion{N}{2}] $\lambda\lambda$6548,6583 and \ion{C}{2} $\lambda$6578 lines have no fine structure and so this correction is not required.  These broadening contributions were assumed to be Gaussian in shape and subtracted in quadrature from the observed line width to obtain the intrinsic line width for each line measurement \citep[for details, see][]{richeretal2008}.  

The \ion{He}{2} $\lambda$6560 line is an exception to the above procedure, as we did not correct it for fine structure broadening.  Qualitatively, the fine structure of this line resembles that of \ion{He}{2} $\lambda$4686 line \citep[e.g.,][]{cleggetal1999}, and is resolved at our spectral resolution.  Due to the weakness of \ion{He}{2} $\lambda$6560, in most cases we would have only measured the line's strongest sub-structure \citep[the sum of the 15 redder components spanning a total velocity range of 7.2 km\,s$^{-1}$: NIST database;][]{ralchenkoetal2011}.  Finally, this sub-structure is narrower than the thermal component.  Hence, our \ion{He}{2} $\lambda$6560 line widths will be slightly over-estimated, but this will not affect our results.  

For the line widths measured from fitting two Gaussian components, no correction was made to account for broadening of any kind since broadening does not affect the central positions of the Gaussian components.  

\begin{figure}[t]
\includegraphics[width=\columnwidth]{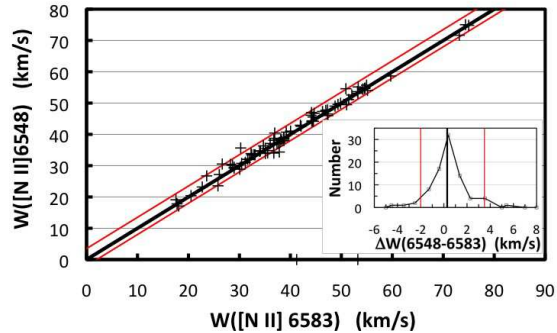}
\caption{The correlation between the line widths (FWHM) measured for the [\ion{N}{2}] $\lambda\lambda$6548,6583 lines is excellent.  The crosses indicate the values measured from individual spectra.  The bold line indicates equality, while the two thin red lines, at $-2.0$ and $+3.5$ km\,s$^{-1}$, bound 90\% of the dispersion about equality.  The inset presents a histogram of the dispersion about equality, i.e., $\mathrm{FWHM}(\lambda 6548) = \mathrm{FWHM}(\lambda 6583)$; the black vertical line is the median value, $+0.29$ km\,s$^{-1}$, and the red lines indicate the 5 and 95 percentile values.  }
\label{fig_comp_n2}
\end{figure}

All of the line widths reported for each spectrum were measured in the same way, using either one (69 cases) or two (14 cases) Gaussian components.  For the 7 objects with two slit positions, a single method was used for both spectra.  Our justification for using two measures of the line width is to use the measure that is most representative of the kinematics of the emission lines in each spectrum.  We always compare the line widths of different emission lines measured by the same method, but never a single emission line measured using the two methods.  In the discussion and figures that follow, we refer only to the intrinsic line widths, corrected for broadening when relevant, and denote these intrinsic line widths by $W$ in Table \ref{table_velocities}.  

We estimate the uncertainties inherent in our measurements of line widths (FWHM) by comparing the line widths of the [\ion{N}{2}] $\lambda\lambda$6548,6583 lines.  These lines arise from the same upper level and so necessarily arise from the same volume within each nebula in our sample.  The two lines should therefore have the same line width.  In Figure \ref{fig_comp_n2}, we present the correlation of the line widths measured for the [\ion{N}{2}] $\lambda\lambda$6548,6583 lines, which is excellent.  The dispersion about the line of equal values has wings that are more extended than a Gaussian, but 90\% of the dispersion is contained within the range from $-2.0$ to $+3.5$ km\,s$^{-1}$ about equality.  Hence, we adopt this velocity uncertainty as the uncertainty in line widths due to the combined effects of our measurement process and astrophysical origins.

\begin{deluxetable}{lc|ccc|ccc|ccc|ccc|ccc|c}
\tabletypesize{\scriptsize}
\rotate
\tablewidth{595pt}
\tablecaption{Line Widths and Radial Velocities}
\tablehead{
\colhead{object$^a$} & \colhead{PN G} & \multicolumn{3}{c}{6548} & \multicolumn{3}{c}{6560} & \multicolumn{3}{c}{6563} & \multicolumn{3}{c}{6578} & \multicolumn{3}{c}{6583} & \colhead{G}\\
\colhead{} & \colhead{} & \colhead{$W_{obs}$} & \colhead{$W$} & \colhead{$V_{rad}$} & \colhead{$W_{obs}$} & \colhead{$W$} & \colhead{$V_{rad}$} & \colhead{$W_{obs}$} & \colhead{$W$} & \colhead{$V_{rad}$} & \colhead{$W_{obs}$} & \colhead{$W$} & \colhead{$V_{rad}$} & \colhead{$W_{obs}$} & \colhead{$W$} & \colhead{$V_{rad}$} & \colhead{} \\
\colhead{(1)} & \colhead{(2)} & \colhead{(3)} & \colhead{(4)} & \colhead{(5)} & \colhead{(6)} & \colhead{(7)} & \colhead{(8)} & \colhead{(9)} & \colhead{(10)} & \colhead{(11)} & \colhead{(12)} & \colhead{(13)} & \colhead{(14)} & \colhead{(15)} & \colhead{(16)} & \colhead{(17)} & \colhead{(18)} }
\startdata
Bl 2-1      & 104.1+01.0 & 47   & 46   & -104 &      &      &      & 46   & 39   & -104 & 48   & 47   & -110 & 46   & 44   & -106 & 1 \\
Cn 1-5      & 002.2-09.4 & 45   & 43   & -29  &      &      &      & 47   & 40   & -30  & 65   & 64   & -32  & 44   & 42   & -30  & 1 \\
H 1-11      & 002.6+08.1 & 48   & 46   & 11   & 36   & 32   & 12   & 42   & 33   & 5    & 40   & 38   & 15   & 46   & 44   & 6    & 1 \\
H 1-18      & 357.6+02.6 & 41   & 39   & -231 & 38   & 35   & -228 & 37   & 27   & -230 & 34   & 31   & -234 & 40   & 38   & -232 & 1 \\
H 1-24      & 004.6+06.0 & 45   & 43   & 162  &      &      &      & 40   & 31   & 160  & 41   & 38   & 162  & 44   & 42   & 161  & 1 \\
H 1-27      & 005.0+04.4 &      &      &      &      &      &      & 47   & 39   & 11   & 35   & 32   & 2    & 55   & 53   & 5    & 1 \\
H 1-33      & 355.7-03.0 & 33   & 31   & -121 & 17   & 4    & -118 & 36   & 25   & -122 & 27   & 24   & -126 & 33   & 30   & -122 & 1 \\
H 1-54      & 002.1-04.2 & 49   & 48   & -111 &      &      &      & 41   & 32   & -114 & 24   & 20   & -114 & 49   & 47   & -111 & 1 \\
H 1-56      & 001.7-04.6 & 49   & 47   & -91  &      &      &      & 38   & 28   & -93  & 39   & 36   & -89  & 46   & 44   & -92  & 1 \\
Hb 12(a)    & 111.8-02.8 & 39   & 37   & -24  &      &      &      & 40   & 31   & -20  & 27   & 23   & -21  & 38   & 35   & -24  & 1 \\
He 2-436    & 004.8-22.7 & 43   & 40   & 118  &      &      &      & 37   & 27   & 116  & 41   & 39   & 112  & 39   & 37   & 113  & 1 \\
He 2-447    & 057.9-01.5 & 36   & 33   & 35   &      &      &      & 34   & 22   & 34   & 21   & 16   & 34   & 35   & 32   & 33   & 1 \\
Hf 2-2(b)   & 005.1-08.9 & 37   & 34   & 85   &      &      &      & 45   & 38   & 82   & 40   & 38   & 80   & 39   & 37   & 82   & 1 \\
Hu 1-1      & 119.6-06.7 & 46   & 44   & -46  & 30   & 25   & -44  & 41   & 33   & -46  & 32   & 29   & -48  & 46   & 44   & -47  & 1 \\
IC 2149(c)  & 166.1+10.4 & 32   & 29   & -52  &      &      &      & 34   & 22   & -49  & 19   & 14   & -46  & 32   & 29   & -54  & 1 \\
IC 2149(d)  & 166.1+10.4 & 38   & 36   & -64  &      &      &      & 36   & 25   & -53  & 19   & 14   & -48  & 33   & 30   & -66  & 1 \\
IC 418(c)   & 215.2-24.2 & 30   & 27   & 77   &      &      &      & 31   & 18   & 75   & 18   & 12   & 74   & 29   & 26   & 76   & 1 \\
IC 4593(j)  & 025.3+40.8 & 35   & 32   & 22   &      &      &      & 35   & 24   & 21   & 22   & 17   & 19   & 33   & 31   & 20   & 1 \\
IC 4593(k)  & 025.3+40.8 & 32   & 29   & 21   &      &      &      & 34   & 22   & 20   & 22   & 17   & 19   & 30   & 26   & 18   & 1 \\
IC 4673     & 003.5-02.4 & 59   & 59   & 4    & 46   & 46   & 4    & 51   & 51   & 1    & 47   & 47   & 0    & 60   & 60   & 3    & 2 \\
J 900       & 194.2+02.5 & 49   & 47   & 21   & 39   & 35   & 15   & 48   & 41   & 16   & 53   & 52   & 25   & 48   & 46   & 20   & 1 \\
K 3-66      & 167.4-09.1 & 55   & 53   & -74  &      &      &      & 50   & 43   & -73  & 37   & 34   & -75  & 55   & 53   & -76  & 1 \\
M 1-14      & 234.9-01.4 & 36   & 34   & 69   &      &      &      & 36   & 25   & 73   & 23   & 18   & 73   & 35   & 33   & 69   & 1 \\
M 1-17      & 228.8+05.3 & 32   & 30   & 55   & 22   & 15   & 58   & 35   & 23   & 54   & 23   & 19   & 56   & 32   & 29   & 53   & 1 \\
M 1-19      & 351.1+04.8 & 51   & 50   & -59  &      &      &      & 40   & 31   & -58  & 34   & 31   & -73  & 51   & 49   & -59  & 1 \\
M 1-20      & 006.1+08.3 & 40   & 38   & 69   &      &      &      & 34   & 23   & 69   & 18   & 13   & 69   & 41   & 38   & 67   & 1 \\
M 1-29      & 359.1-01.7 & 50   & 48   & -45  & 40   & 36   & -41  & 47   & 40   & -43  & 40   & 38   & -24  & 49   & 47   & -46  & 1 \\
M 1-30      & 355.9-04.2 &      &      &      &      &      &      & 40   & 30   & -121 & 32   & 29   & -122 & 43   & 41   & -122 & 1 \\
M 1-31      & 006.4+02.0 & 41   & 39   & 90   &      &      &      & 36   & 26   & 90   & 29   & 25   & 88   & 41   & 39   & 89   & 1 \\
M 1-33      & 013.1+04.1 & 52   & 50   & -51  & 44   & 41   & -44  & 46   & 38   & -50  & 37   & 34   & -50  & 52   & 50   & -54  & 1 \\
M 1-35      & 003.9-02.3 & 56   & 54   & 97   & 44   & 41   & 99   & 47   & 39   & 95   & 44   & 42   & 93   & 56   & 54   & 94   & 1 \\
M 1-37      & 002.6-03.4 & 35   & 32   & 215  &      &      &      & 39   & 30   & 214  & 28   & 25   & 210  & 34   & 32   & 213  & 1 \\
M 1-42(b)   & 002.7-04.8 & 39   & 36   & -91  & 32   & 28   & -93  & 48   & 40   & -91  & 44   & 42   & -94  & 37   & 35   & -92  & 1 \\
M 1-42(c)   & 002.7-04.8 & 40   & 38   & -104 & 39   & 35   & -97  & 48   & 41   & -102 & 49   & 47   & -103 & 38   & 36   & -105 & 1 \\
M 1-46      & 016.4-01.9 & 43   & 41   & 35   &      &      &      & 41   & 31   & 31   & 36   & 33   & 31   & 41   & 39   & 33   & 1 \\
M 1-48      & 013.4-03.9 & 48   & 47   & 147  & 18   & 8    & 145  & 38   & 28   & 146  & 25   & 21   & 145  & 49   & 47   & 145  & 1 \\
M 1-5       & 184.0-02.1 & 27   & 23   & 56   &      &      &      & 36   & 26   & 56   & 20   & 15   & 55   & 26   & 23   & 55   & 1 \\
M 1-59      & 023.9-02.3 & 32   & 29   & 78   & 38   & 35   & 79   & 43   & 35   & 76   & 42   & 40   & 78   & 31   & 28   & 76   & 1 \\
M 1-6       & 211.2-03.5 & 22   & 18   & 75   &      &      &      & 30   & 15   & 74   & 15   & 8    & 75   & 22   & 18   & 74   & 1 \\
M 1-61      & 019.4-05.3 & 46   & 44   & 3    &      &      &      & 55   & 49   & 5    & 35   & 32   & 2    & 46   & 44   & 2    & 1 \\
M 1-65      & 043.1+03.8 & 29   & 26   & 26   &      &      &      & 33   & 20   & 25   & 19   & 13   & 23   & 26   & 23   & 24   & 1 \\
M 1-8       & 210.3+01.9 & 46   & 44   & 46   & 34   & 30   & 51   & 45   & 37   & 47   & 37   & 34   & 66   & 46   & 44   & 44   & 1 \\
M 1-9       & 212.0+04.3 & 39   & 37   & 123  &      &      &      & 34   & 22   & 123  & 22   & 18   & 124  & 39   & 37   & 122  & 1 \\
M 2-14      & 003.6+03.1 & 37   & 35   & -48  &      &      &      & 35   & 24   & -48  & 25   & 21   & -50  & 37   & 34   & -49  & 1 \\
M 2-16      & 357.4-03.2 & 56   & 55   & 90   & 48   & 45   & 91   & 51   & 44   & 90   & 31   & 28   & 81   & 56   & 55   & 89   & 1 \\
M 2-19      & 000.2-01.9 & 37   & 35   & -20  &      &      &      & 34   & 23   & -25  & 20   & 15   & -29  & 36   & 34   & -23  & 1 \\
M 2-27      & 359.9-04.5 & 39   & 37   & 147  &      &      &      & 44   & 36   & 148  & 41   & 38   & 149  & 40   & 38   & 145  & 1 \\
M 2-29      & 004.0-03.0 & 22   & 17   & -122 &      &      &      & 36   & 25   & -111 & 29   & 25   & -116 & 22   & 18   & -122 & 1 \\
M 2-30      & 003.7-04.6 & 40   & 37   & 161  & 50   & 47   & 157  & 54   & 48   & 157  & 25   & 21   & 161  & 40   & 38   & 159  & 1 \\
M 2-31      & 006.0-03.6 & 76   & 75   & 152  &      &      &      & 55   & 48   & 155  & 26   & 22   & 150  & 76   & 75   & 150  & 1 \\
M 2-33      & 002.0-06.2 & 24   & 20   & -100 &      &      &      & 32   & 19   & -101 & 23   & 18   & -103 & 24   & 20   & -100 & 1 \\
M 2-36      & 003.2-06.2 & 53   & 52   & 58   & 21   & 14   & 57   & 40   & 30   & 55   & 29   & 26   & 54   & 53   & 51   & 57   & 1 \\
M 2-39      & 008.1-04.7 & 23   & 19   & 84   &      &      &      & 47   & 40   & 80   & 32   & 30   & 75   & 22   & 18   & 83   & 1 \\
M 2-49      & 095.1-02.0 & 53   & 51   & -108 & 29   & 24   & -108 & 40   & 31   & -110 & 29   & 26   & -109 & 53   & 51   & -110 & 1 \\
M 2-8       & 352.1+05.1 & 35   & 32   & 23   & 54   & 51   & 21   & 45   & 36   & 22   & 36   & 34   & 19   & 35   & 32   & 21   & 1 \\
M 3-12      & 005.2+05.6 & 55   & 54   & 38   & 50   & 48   & 34   & 58   & 52   & 32   & 61   & 59   & 34   & 55   & 53   & 35   & 1 \\
M 3-14      & 355.4-02.4 & 56   & 55   & -84  & 52   & 50   & -80  & 53   & 46   & -86  & 37   & 34   & -88  & 56   & 55   & -86  & 1 \\
M 3-17      & 359.3-03.1 & 39   & 37   & -46  &      &      &      & 38   & 28   & -46  & 27   & 23   & -44  & 38   & 36   & -47  & 1 \\
M 3-32      & 009.4-09.8 & 73   & 72   & 73   & 44   & 42   & 75   & 54   & 47   & 74   & 40   & 37   & 73   & 74   & 73   & 69   & 1 \\
M 3-6       & 253.9+05.7 & 43   & 41   & 48   &      &      &      & 38   & 28   & 43   & 28   & 25   & 42   & 42   & 40   & 47   & 1 \\
Me 2-2      & 100.0-08.7 & 34   & 31   & -139 &      &      &      & 33   & 20   & -140 & 23   & 19   & -142 & 34   & 31   & -142 & 1 \\
NGC 1535(a) & 206.4-40.5 & 47   & 47   & 18   & 32   & 32   & 20   & 39   & 39   & 18   & 45   & 45   & 18   & 45   & 45   & 16   & 2 \\
NGC 1535(b) & 206.4-40.5 & 38   & 38   & 20   & 18   & 18   & 21   & 33   & 33   & 22   & 31   & 31   & 21   & 39   & 39   & 19   & 2 \\
NGC 3242(j) & 261.1+32.0 & 38   & 38   & -14  & 32   & 32   & -13  & 37   & 37   & -15  & 37   & 37   & -15  & 39   & 39   & -17  & 2 \\
NGC 3242(k) & 261.1+32.0 & 29   & 29   & -14  & 24   & 24   & -12  & 29   & 29   & -13  & 29   & 29   & -14  & 30   & 30   & -15  & 2 \\
NGC 40      & 120.0+09.8 & 49   & 49   & -14  &      &      &      & 48   & 48   & -15  & 46   & 46   & -18  & 49   & 49   & -15  & 2 \\
NGC 6153    & 341.8+05.4 & 49   & 49   & 29   & 35   & 35   & 30   & 44   & 44   & 27   & 42   & 42   & 25   & 51   & 51   & 25   & 2 \\
NGC 6439    & 011.0+05.8 & 57   & 55   & -102 & 44   & 41   & -91  & 50   & 43   & -95  & 41   & 38   & -93  & 55   & 53   & -104 & 1 \\
NGC 6543(a) & 096.4+29.9 & 39   & 39   & -65  &      &      &      & 32   & 32   & -68  & 26   & 26   & -67  & 39   & 39   & -69  & 2 \\
NGC 6543(b) & 096.4+29.9 & 34   & 34   & -67  &      &      &      & 33   & 33   & -67  & 30   & 30   & -67  & 35   & 35   & -68  & 2 \\
NGC 6567    & 011.7-00.6 & 55   & 55   & 114  &      &      &      & 43   & 43   & 122  & 29   & 29   & 116  & 54   & 54   & 113  & 2 \\
NGC 6629    & 009.4-05.5 & 37   & 34   & 10   &      &      &      & 39   & 29   & 10   & 23   & 18   & 9    & 38   & 36   & 9    & 1 \\
NGC 6778    & 034.5-06.7 & 76   & 75   & 109  & 26   & 21   & 110  & 51   & 44   & 109  & 25   & 21   & 110  & 76   & 74   & 106  & 1 \\
NGC 6826    & 083.5+12.7 & 48   & 46   & 0    &      &      &      & 52   & 46   & -1   & 30   & 26   & -3   & 49   & 47   & -2   & 1 \\
NGC 7009    & 037.7-34.5 & 37   & 37   & -17  & 30   & 30   & -15  & 34   & 34   & -18  & 32   & 32   & -18  & 37   & 37   & -19  & 2 \\
NGC 7354    & 107.8+02.2 & 54   & 54   & -46  & 43   & 43   & -46  & 48   & 48   & -48  & 48   & 48   & -49  & 55   & 55   & -49  & 2 \\
NGC 7662(c) & 106.5-17.6 & 34   & 34   & -13  & 32   & 32   & -8   & 37   & 37   & -12  & 29   & 29   & -18  & 38   & 38   & -15  & 2 \\
NGC 7662(d) & 106.5-17.6 & 35   & 35   & 6    & 36   & 36   & 4    & 38   & 38   & 2    & 40   & 40   & 2    & 36   & 36   & 5    & 2 \\
PC 12       & 000.1+17.2 & 36   & 33   & -62  &      &      &      & 38   & 28   & -62  & 39   & 36   & -78  & 36   & 33   & -62  & 1 \\
PC 24       & 066.9-05.2 & 57   & 55   & -47  & 45   & 42   & -45  & 49   & 41   & -48  & 40   & 38   & -52  & 56   & 55   & -49  & 1 \\
Ps 1        & 065.0-27.3 & 27   & 24   & -93  &      &      &      & 41   & 32   & -96  & 26   & 22   & -94  & 29   & 26   & -94  & 1 \\
Sp 4-1      & 068.7+14.8 & 37   & 35   & -83  &      &      &      & 39   & 30   & -83  & 31   & 28   & -83  & 37   & 35   & -84  & 1 \\
Vy 1-1      & 118.0-08.6 & 33   & 30   & -38  &      &      &      & 34   & 23   & -40  & 24   & 20   & -39  & 29   & 26   & -41  & 1 \\
\enddata
\label{table_velocities}

\tablenotetext{a}{The letter in parentheses indicates the slit position from the online SPM Catalogue, http://kincatpn.astrosen.unam.mx/.}
\tablecomments{The line widths have uncertainties of $\pm 2-3$ km\,s$^{-1}$ (see Figure \ref{fig_comp_n2}).  Radial velocities have lower formal uncertainties, but may suffer from systematic uncertainties due to wavelength calibration of order $\pm 2$ km\,s$^{-1}$ (see Figure \ref{fig_vrad}).}
\end{deluxetable}

\begin{deluxetable}{lcccccccccc}
\tabletypesize{\footnotesize}
\tablewidth{0pt}
\tablecaption{Nebular Sizes}
\tablehead{
\colhead{object$^a$} & \colhead{PN G} & \multicolumn{4}{l}{Diameters$^b$  (in arcseconds)} & \colhead{Q$^c$} & \colhead{$P_s$} & \colhead{$R$} & \colhead{ADF$(\mathrm O^{2+})$} & \colhead{Ref.} \\
\colhead{} & \colhead{} & \colhead{6560} & \colhead{6563} & \colhead{6578} & \colhead{6583} & \colhead{} & \colhead{\arcsec$/\mathrm{pix}$} & \colhead{\AA$/\mathrm{pix}$} & & \\
\colhead{(1)} & \colhead{(2)} & \colhead{(3)} & \colhead{(4)} & \colhead{(5)} & \colhead{(6)} & \colhead{(7)} & \colhead{(8)} & \colhead{(9)} & \colhead{(10)} & \colhead{(11)}}
\startdata
Bl 2-1      & 104.1+01.0 &            &  5.8 &    4 &  6.4 &       0 &   0.624 &   0.1011 &           &     \\
Cn 1-5      & 002.2-09.4 &            &  8.6 &    8 &  8.8 &       0 &   0.624 &   0.1007 &      1.8  & K,M \\
H 1-11      & 002.6+08.1 &        3.1 &  7.7 &    9 &  7.8 &       0 &   0.624 &   0.1003 &           &     \\
H 1-18      & 357.6+02.6 &        2.5 &  4.9 &    4 &  5.6 &       0 &   0.624 &   0.1007 &           &     \\
H 1-24      & 004.6+06.0 &            &  7.6 &    6 &  7.7 &       0 &   0.624 &   0.1003 &           &     \\
H 1-27      & 005.0+04.4 &            &  3.8 &  3.7 &  4.1 &       1 &   0.624 &   0.0994 &           &     \\
H 1-33      & 355.7-03.0 &          5 &  6.9 &    6 &  6.7 &       0 &   0.624 &   0.1006 &           &     \\
H 1-54      & 002.1-04.2 &            &  5.4 &    4 &  5.7 &       0 &   0.624 &   0.0998 &      2.5  &   K \\
H 1-56      & 001.7-04.6 &            &  6.1 &    4 &  8.2 &       0 &   0.624 &   0.0999 &           &     \\
Hb 12(a)    & 111.8-02.8 &            &  5.8 &  3.2 &  8.6 &       1 &   0.312 &   0.0503 &           &     \\
He 2-436    & 004.8-22.7 &            &  3.9 &    4 &  3.9 &       0 &   0.624 &   0.1006 &     16.2  &   L \\
He 2-447    & 057.9-01.5 &            &  5.6 &  4.6 &  6.0 &       1 &   0.624 &   0.1009 &           &     \\
Hf 2-2(b)   & 005.1-08.9 &            & 23.2 & 20.2 & 23.8 &       1 &   0.624 &   0.1003 &     70.0  &   J \\
Hu 1-1      & 119.6-06.7 &        9.4 & 10.8 &    9 & 12.5 &       0 &   0.624 &   0.1004 &      3.0  &   I \\
IC 2149(c)  & 166.1+10.4 &            & 13.4 &    6 &  9.9 &       0 &   0.624 &   0.1021 &           &     \\
IC 2149(d)  & 166.1+10.4 &            & 14.0 &   13 & 13.8 &       1 &   0.624 &   0.1021 &           &     \\
IC 418(c)   & 215.2-24.2 &            & 16.2 & 15.6 & 16.8 &       1 &   0.624 &   0.1005 &      2.0  &   E \\
IC 4593(j)  & 025.3+40.8 &            & 14.4 & 15.3 & 16.3 &       1 &    0.35 &   0.0576 &      3.6  &   H \\
IC 4593(k)  & 025.3+40.8 &            & 14.1 & 14.3 & 17.1 &       1 &    0.35 &   0.0572 &      3.6  &   H \\
IC 4673     & 003.5-02.4 &       30.3 & 26.3 & 22.8 & 29.0 &       1 &   0.624 &   0.0854 &           &     \\
J 900       & 194.2+02.5 &        9.4 &  9.9 &   10 & 10.6 &       0 &   0.624 &   0.0995 &           &     \\
K 3-66      & 167.4-09.1 &            &  7.6 &    6 &  7.7 &       0 &   0.624 &   0.1004 &           &     \\
M 1-14      & 234.9-01.4 &            &  6.9 &    5 &  7.4 &       0 &   0.624 &   0.1003 &           &     \\
M 1-17      & 228.8+05.3 &        5.1 &  5.6 &  4.9 &  6.1 &       1 &   0.624 &   0.1008 &           &     \\
M 1-19      & 351.1+04.8 &            &  6.0 &    4 &  6.8 &       0 &   0.624 &   0.1005 &           &     \\
M 1-20      & 006.1+08.3 &            &  5.2 &    4 &  5.2 &       0 &   0.624 &   0.1007 &      1.4  &   K \\
M 1-29      & 359.1-01.7 &        7.6 &  9.7 &   13 & 10.4 &       0 &   0.624 &   0.1007 &      2.9  &   K \\
M 1-30      & 355.9-04.2 &            &  6.5 &    6 &  7.5 &       0 &   0.624 &   0.1003 &      2.1  &   M \\
M 1-31      & 006.4+02.0 &            &  4.7 &    4 &  5.2 &       0 &   0.624 &   0.1006 &           &     \\
M 1-33      & 013.1+04.1 &        5.4 &  5.9 &    6 &  6.6 &       1 &   0.624 &   0.0992 &           &     \\
M 1-35      & 003.9-02.3 &        5.0 &  7.3 &  5.7 &  7.9 &       1 &   0.624 &   0.1001 &           &     \\
M 1-37      & 002.6-03.4 &            &  6.3 &    4 &  6.4 &       0 &   0.624 &   0.0993 &           &     \\
M 1-42(b)   & 002.7-04.8 &        4.4 &  7.1 &  5.7 &  8.5 &       1 &    0.35 &   0.0999 &     14.5  &   C \\
M 1-42(c)   & 002.7-04.8 &        9.3 & 13.2 &   13 & 15.0 &       0 &    0.35 &   0.0574 &     14.5  &   C \\
M 1-46      & 016.4-01.9 &            & 13.0 &   11 & 13.5 &       0 &   0.624 &   0.0999 &           &     \\
M 1-48      & 013.4-03.9 &        6.5 &  7.0 &    4 &  7.9 &       0 &   0.624 &   0.1006 &           &     \\
M 1-5       & 184.0-02.1 &            &  6.1 &  5.9 &  6.1 &       1 &   0.624 &   0.1007 &           &     \\
M 1-59      & 023.9-02.3 &        5.7 &  7.2 &    8 &  8.9 &       1 &   0.525 &   0.0852 &           &     \\
M 1-6       & 211.2-03.5 &            &  6.1 &    4 &  6.6 &       0 &   0.624 &   0.1002 &           &     \\
M 1-61      & 019.4-05.3 &            &  5.5 &  5.1 &  5.1 &       1 &   0.525 &   0.0841 &      1.8  & K,M \\
M 1-65      & 043.1+03.8 &            &  6.0 &    5 &  6.2 &       0 &   0.525 &   0.0862 &           &     \\
M 1-8       & 210.3+01.9 &       14.5 & 19.0 &   20 & 21.8 &       0 &   0.624 &   0.1012 &           &     \\
M 1-9       & 212.0+04.3 &            &  5.7 &    5 &  6.2 &       0 &   0.624 &   0.1005 &           &     \\
M 2-14      & 003.6+03.1 &            &  5.4 &    4 &  6.2 &       0 &   0.624 &   0.1003 &           &     \\
M 2-16      & 357.4-03.2 &        5.2 &  5.1 &    4 &  5.9 &       0 &   0.624 &   0.1007 &           &     \\
M 2-19      & 000.2-01.9 &            &  8.6 &    8 & 10.7 &       0 &   0.624 &   0.1000 &           &     \\
M 2-27      & 359.9-04.5 &            &  5.1 &  4.4 &  5.8 &       1 &   0.624 &   0.1007 &      2.2  &   K \\
M 2-29      & 004.0-03.0 &            &  7.3 &    5 &  7.4 &       0 &   0.624 &   0.1007 &           &     \\
M 2-30      & 003.7-04.6 &        5.8 &  6.3 &    4 &  6.6 &       0 &   0.624 &   0.1008 &           &     \\
M 2-31      & 006.0-03.6 &            &  6.1 &    4 &  5.6 &       0 &   0.624 &   0.1007 &           &     \\
M 2-33      & 002.0-06.2 &            &  7.1 &    4 &  8.0 &       0 &   0.624 &   0.0999 &      1.4  &   K \\
M 2-36      & 003.2-06.2 &        6.6 &  7.4 &  7.1 &  9.0 &       1 &    0.35 &   0.0577 &      6.2  &   C \\
M 2-39      & 008.1-04.7 &            &  4.7 &    2 &  4.2 &       0 &   0.624 &   0.1004 &      3.5  &   J \\
M 2-49      & 095.1-02.0 &        6.1 &  6.2 &    4 &  7.1 &       0 &   0.624 &   0.1012 &           &     \\
M 2-8       & 352.1+05.1 &        6.1 &  5.8 &    4 &  5.8 &       0 &   0.624 &   0.1003 &           &     \\
M 3-12      & 005.2+05.6 &        8.1 &  8.9 &    9 &  9.7 &       1 &   0.624 &   0.0998 &           &     \\
M 3-14      & 355.4-02.4 &          6 &  8.2 &    5 & 10.5 &       0 &   0.624 &   0.1006 &           &     \\
M 3-17      & 359.3-03.1 &            &  6.2 &  5.9 &  6.6 &       1 &   0.624 &   0.1003 &           &     \\
M 3-32      & 009.4-09.8 &        6.9 &  8.0 &  8.7 &  7.2 &       1 &   0.624 &   0.1006 &      1.8  &   K \\
M 3-6       & 253.9+05.7 &            & 14.0 &    9 & 15.7 &       0 &   0.624 &   0.1007 &           &     \\
Me 2-2      & 100.0-08.7 &            &  5.9 &  6.7 &  6.2 &       1 &   0.624 &   0.1006 &      2.7  &   I \\
NGC 1535(a) & 206.4-40.5 &       18.9 & 34.3 &   20 & 19.8 &       0 &    0.48 &   0.0799 &           &     \\
NGC 1535(b) & 206.4-40.5 &       20.2 & 34.1 & 20.4 & 21.7 &       1 &    0.48 &   0.0799 &           &     \\
NGC 3242(j) & 261.1+32.0 &       23.9 & 37.2 &   26 & 43.1 &       0 &   0.624 &   0.1002 &      2.5  &   F \\
NGC 3242(k) & 261.1+32.0 &       21.2 & 40.1 &   27 & 40.7 &       0 &   0.624 &   0.1002 &      2.5  &   F \\
NGC 40      & 120.0+09.8 &            & 54.3 &   42 & 54.9 &       0 &   0.585 &   0.0629 &     18.2  &   D \\
NGC 6153    & 341.8+05.4 &       28.9 & 32.1 & 33.4 & 32.8 &       1 &    0.35 &   0.0573 &      9.3  &   B \\
NGC 6439    & 011.0+05.8 &        6.2 &  6.9 &  5.2 &  7.8 &       1 &   0.624 &   0.1001 &      6.2  &   K \\
NGC 6543(a) & 096.4+29.9 &            & 21.8 & 19.2 & 21.5 &       1 &   0.312 &   0.0573 &      2.8  &   G \\
NGC 6543(b) & 096.4+29.9 &            & 18.3 &   16 & 18.3 &       0 &   0.312 &   0.0573 &      2.8  &   G \\
NGC 6567    & 011.7-00.6 &        9.6 &  9.7 & 10.8 & 11.2 &       1 &   0.624 &   0.1001 &      2.2  &   K \\
NGC 6629    & 009.4-05.5 &            & 17.8 & 15.5 & 18.6 &       1 &   0.624 &   0.1007 &           &     \\
NGC 6778    & 034.5-06.7 &       17.7 & 21.4 & 23.2 & 15.0 &       1 &   0.624 &   0.0996 &     17.9  &   O \\
NGC 6826    & 083.5+12.7 &            & 31.3 & 35.9 & 31.4 &       1 &   0.624 &   0.1003 &      1.9  &   D \\
NGC 7009    & 037.7-34.5 &       24.3 & 29.0 & 20.5 & 52.7 &       1 &   0.624 &   0.0996 &      4.7  &   A \\
NGC 7354    & 107.8+02.2 &       24.8 & 31.6 & 30.5 & 32.6 &       1 &   0.624 &   0.0996 &           &     \\
NGC 7662(c) & 106.5-17.6 &       16.2 & 27.6 &   21 & 31.9 &       1 &   0.624 &   0.0996 &      2.0  &   D \\
NGC 7662(d) & 106.5-17.6 &       18.5 & 27.8 & 15.5 & 32.4 &       1 &   0.624 &   0.0996 &      2.0  &   D \\
PC 12       & 000.1+17.2 &            &  6.6 &    3 &  6.9 &       0 &   0.624 &   0.1005 &           &     \\
PC 24       & 066.9-05.2 &        5.4 &  6.2 &    6 &  6.4 &       1 &   0.525 &   0.0861 &           &     \\
Ps 1        & 065.0-27.3 &            &  7.4 &    7 &  7.5 &       0 &   0.624 &   0.0996 &      3.0  &   N \\
Sp 4-1      & 068.7+14.8 &            &  3.7 &  3.9 &  4.0 &       1 &   0.525 &   0.0860 &      2.9  &   I \\
Vy 1-1      & 118.0-08.6 &            &  9.9 &    7 & 11.6 &       0 &   0.585 &   0.0931 &           &     \\
\enddata
\label{table_diameters}

\tablenotetext{a}{The letter in parentheses indicates the slit position from the online SPM Catalogue,\\ http://kincatpn.astrosen.unam.mx/.}
\tablenotetext{b}{The diameters rounded to integer values have an uncertainty $\pm 1$\arcsec, others have an uncertainty of $\pm 0.5$\arcsec.  }
\tablenotetext{c}{A value of zero indicates a low signal-to-noise C~{\sc ii} $\lambda$6578 spatial line profile.}
\tablecomments{References for ADFs:  A--\citet{liuetal1995}; B--\citet{liuetal2000}; C--\citet{liuetal2001}; D--\citet{liuetal2004}; E--\citet{sharpeeetal2004}; F--\citet{tsamisetal2004}; G--\citet{wessonliu2004}; H--\citet{robertsontessigarnett2005}; I--\citet{wessonetal2005}; J--\citet{liuetal2006}; K--\citet{wangliu2007}; L--\citet{otsukaetal2011}; M--\citet{garciarojasetal2013}; N--\citet{otsukaetal2015}; O--\citet{jonesetal2016}}
\end{deluxetable}

\begin{figure}[t]
\includegraphics[width=\columnwidth]{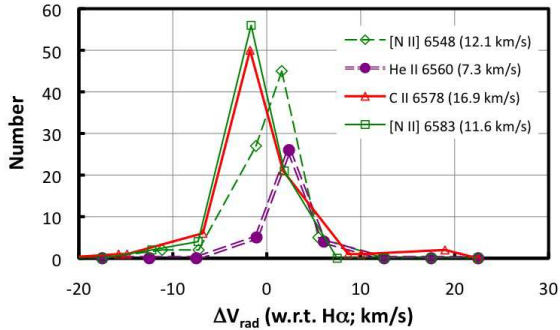}
\caption{We plot the distribution of the difference in radial velocities measured between the other lines and H$\alpha$, i.e., $\Delta V_{rad}(\mathrm{line}) = V_{rad}(\mathrm{line}) - V_{rad}(\mathrm H\alpha)$.  The shapes of the distributions of the difference in radial velocity for the [\ion{N}{2}] $\lambda\lambda$6548,6583, \ion{He}{2} $\lambda$6560, and \ion{C}{2} $\lambda$6578 lines do not differ significantly, indicating that the measured \ion{C}{2} $\lambda$6578 emission is real and not confused with faint artifacts (ghosts, reflections, etc.).   
In the legend, we indicate the velocity interval containing 90\% of the data points for each line.}
\label{fig_vrad}
\end{figure}

The \ion{C}{2} $\lambda$6578 line is often very weak (see Figure \ref{fig_examples}).  In Figure \ref{fig_vrad}, we present the distribution of the difference in radial velocity between the H$\alpha$ line and the lines of [\ion{N}{2}] $\lambda\lambda$6548,6583, \ion{He}{2} $\lambda$6560, and \ion{C}{2} $\lambda$6578.  There are slight offsets between the peaks of the distributions.  Given that \ion{C}{2} $\lambda$6578 is most like [\ion{N}{2}] $\lambda$6583 and that \ion{He}{2} $\lambda$6560 is most like [\ion{N}{2}] $\lambda$6548, i.e., the lines closest in wavelength agree best, the most likely cause of the shifts of the line peaks is that the wavelength solution that was applied (a linear function), is not sufficiently accurate over the whole order.  (Other possible explanations are inconsistent wavelengths or the ionization structure of the nebulae.)  On the other hand, the shape of the distributions is similar for all lines, implying that the feature attributed to \ion{C}{2} $\lambda$6578 does indeed correspond to real emission from this line.  All of the distributions shown here are wider than that in Figure \ref{fig_comp_n2}; the interval containing 90\% of the data is given in brackets in the legend. 

\subsection{Nebular Diameters}

We also measure the sizes of all of the planetary nebulae in the \ion{He}{2} $\lambda$6560, H$\alpha$, \ion{C}{2} $\lambda$6578, and [\ion{N}{2}] $\lambda$6583 lines.  To measure the spatial extent, we integrate the two-dimensional line profiles in the spectral direction to produce one-dimensional spatial profiles, equivalent to what would be observed in spectra of low spectral resolution.  We measure the spatial size as the width where the intensity falls to 10\% of the maximum intensity.  When the signal-to-noise (S/N) of the \ion{He}{2} $\lambda$6560 and \ion{C}{2} $\lambda$6578 lines (2 and 47 cases, respectively) did not allow us to distinguish the 10\% of maximum intensity level from the noise, we instead measured the total spatial extent of the line as observed in the two-dimensional spectrum.  

Table \ref{table_diameters} lists the nebular sizes measured for the planetary nebulae in our sample.  When the S/N is low (the intensity at 10\% of maximum intensity was within the noise), the sizes are rounded to an integer value (in arcseconds).  For these measurements, identified as $Q = 0$ in column 7 of Table \ref{table_diameters}, the formal uncertainty in the measurement is typically about $\pm 1^{\prime\prime}$, though, as we shall see, there are larger systematic uncertainties.  Otherwise ($Q=1$ in column 7), the uncertainty in the measurement is typically $\pm 0.5^{\prime\prime}$.  Column 8 in Table \ref{table_diameters} presents the plate scale for each spectrum.  The last two columns in Table \ref{table_diameters} list abundance discrepancy factors from the literature and their references.  

\section{Results}

\subsection{\ion{C}{2} $\lambda$6578 line widths}\label{sec_c2_line_widths}

For each of the [\ion{N}{2}] $\lambda\lambda$6548,6583, \ion{He}{2} $\lambda$6560 (when present), H$\alpha$, and \ion{C}{2} $\lambda$6578 emission lines, Table \ref{table_velocities} presents the line widths measured in all of our spectra, both as observed and corrected for broadening contributions.  For the line widths based upon two Gaussian components, these two line widths are equal.  Table \ref{table_velocities} also presents the radial velocity for all lines \citep[previously corrected to heliocentric values;][]{lopezetal2012}.  The last column indicates the number of Gaussian components used to measure the line widths reported for each spectrum.  Due to space limitations, the spectral resolution (in \AA$/\mathrm{pix}$) for each spectrum is given in column 9 of Table \ref{table_diameters}.  

\begin{figure}
\includegraphics[width=\columnwidth]{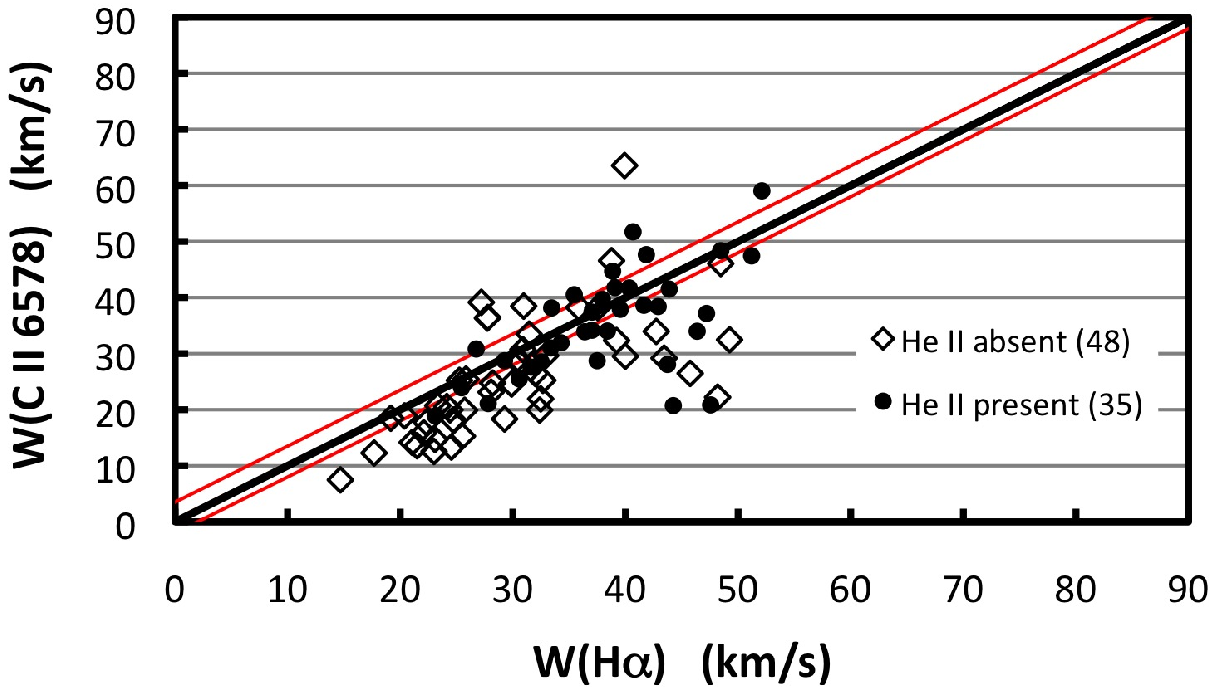}
\includegraphics[width=\columnwidth]{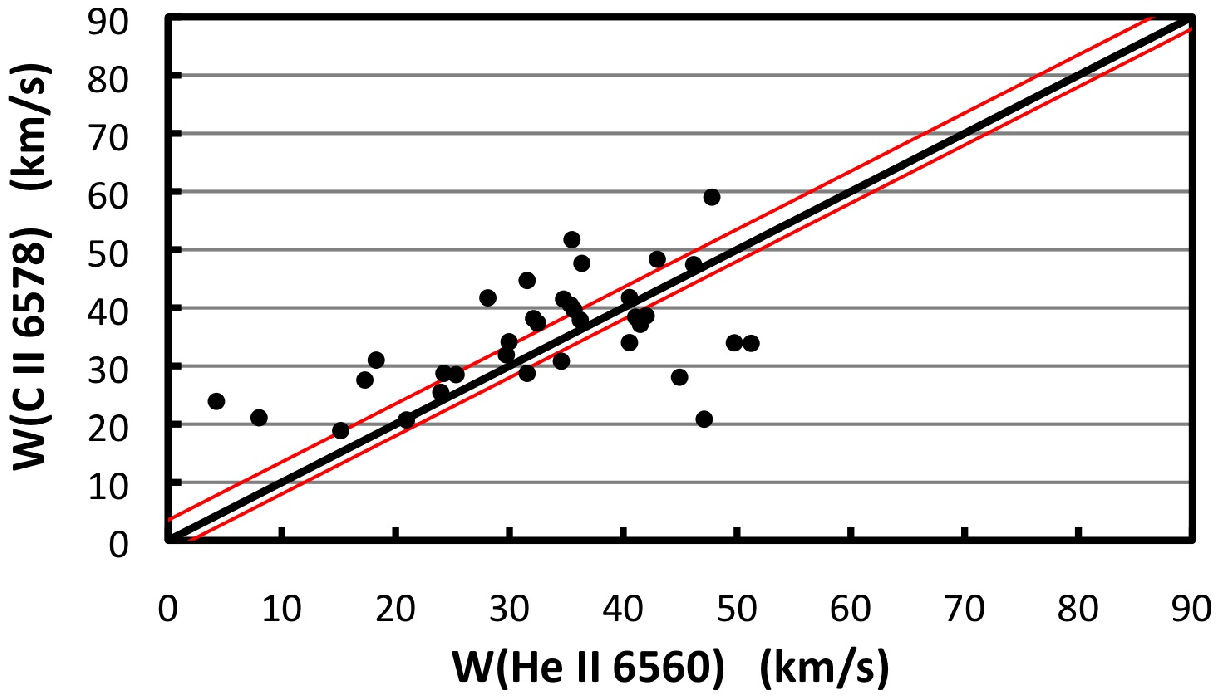}
\includegraphics[width=\columnwidth]{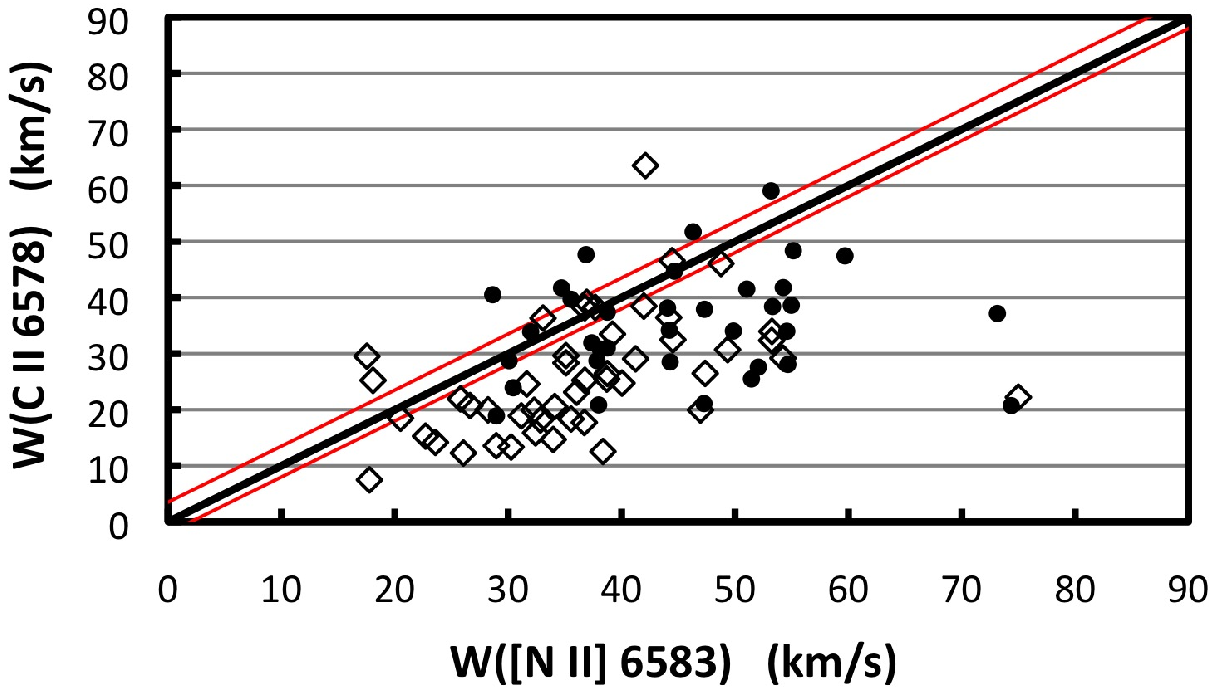}
\caption{We present the correlation of the \ion{C}{2} $\lambda$6578 line width as a function of the line widths of H$\alpha$ (top), \ion{He}{2} $\lambda$6560 (middle), and [\ion{N}{2}] $\lambda$6583 (bottom).  In each panel, the solid black line indicates equal line widths and the two red lines bound the same 90\% interval about equality as in Figure \ref{fig_comp_n2}.  Typically, the \ion{C}{2} $\lambda$6578 line is narrower than the H$\alpha$ and [\ion{N}{2}] lines, but broader than the \ion{He}{2} $\lambda$6560 line.  The number in parentheses indicates the number of objects in each group.  }
\label{fig_fwhm_correlations}
\end{figure}

In Figure \ref{fig_fwhm_correlations}, we present the correlation of the \ion{C}{2} $\lambda$6578 line width as a function of the line widths of the H$\alpha$, \ion{He}{2} $\lambda$6560, and [\ion{N}{2}] $\lambda$6583 lines.  Compared to the dispersion observed in Figure \ref{fig_comp_n2}, it is clear that the dispersion in the three panels of Figure \ref{fig_fwhm_correlations} is substantially greater, implying real differences in the line widths.  Typically, the width of the \ion{C}{2} $\lambda$6578 line is narrower than H$\alpha$, substantially narrower than the [\ion{N}{2}] lines, but broader than the \ion{He}{2} $\lambda$6560 line.  

\begin{figure}[t]
\includegraphics[width=\columnwidth]{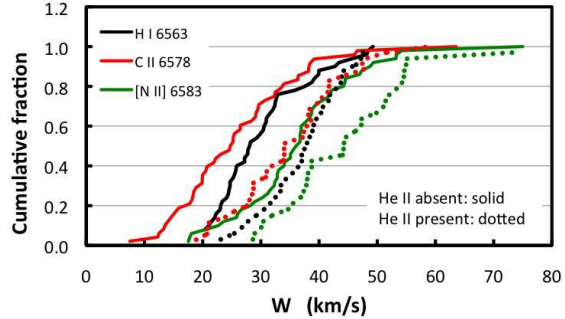}
\caption{The cumulative distributions of the line widths of H$\alpha$, \ion{C}{2} $\lambda$6578, and [\ion{N}{2}] $\lambda$6583 when \ion{He}{2} $\lambda$6560 is absent (solid lines) or present (dotted lines) reflect the acceleration of the nebular shell as a result of the central star evolution \citep{richeretal2008}.  Nebulae with \ion{He}{2} $\lambda$6560 emission have hotter, more evolved central stars and their nebular shells expand faster.  }
\label{fig_fwhm_evol}
\end{figure}

Figure \ref{fig_fwhm_evol} presents the cumulative distributions of the line widths in H$\alpha$, [\ion{N}{2}] $\lambda$6583, and \ion{C}{2} $\lambda$6578 when the \ion{He}{2} $\lambda$6560 line is absent and present.  (These distributions indicate the fraction of objects with line widths less than the line width given by the abscissa value.)  The H$\alpha$, [\ion{N}{2}] $\lambda$6583, and \ion{C}{2} $\lambda$6578 line widths are all shifted to larger values when the \ion{He}{2} $\lambda$6560 line is present.  We compare these distributions using the non-parametric U-test \citep{walljenkins2003}, adopting the criterion that the distributions are statistically distinct if the probability that they arise randomly by chance from the same progenitor population is less than 0.02 (2\%).  By this standard, it is very unlikely that the distributions of the line widths when \ion{He}{2} $\lambda$6560 is present and absent arise from the same parent population, the probabilities being $4.1\times 10^{-4}$, $1.7\times 10^{-6}$, and $2.6\times 10^{-6}$ for the [\ion{N}{2}] $\lambda$6583, \ion{C}{2} $\lambda$6578, and H$\alpha$ lines, respectively.  The nebulae that present \ion{He}{2} $\lambda$6560 emission contain hotter, more evolved central stars.  Thus, as was previously found for a homogeneous sample of planetary nebulae in the bulge of the Milky Way \citep{richeretal2008}, the nebular shells for the more evolved objects in the present sample are also accelerated as the central star evolves.  For both evolutionary stages, the same order of the line widths holds:  \ion{C}{2} $\lambda$6578 is narrowest, H$\alpha$ is intermediate, and [\ion{N}{2}] $\lambda$6583 is widest.  

\begin{figure}
\includegraphics[width=\columnwidth]{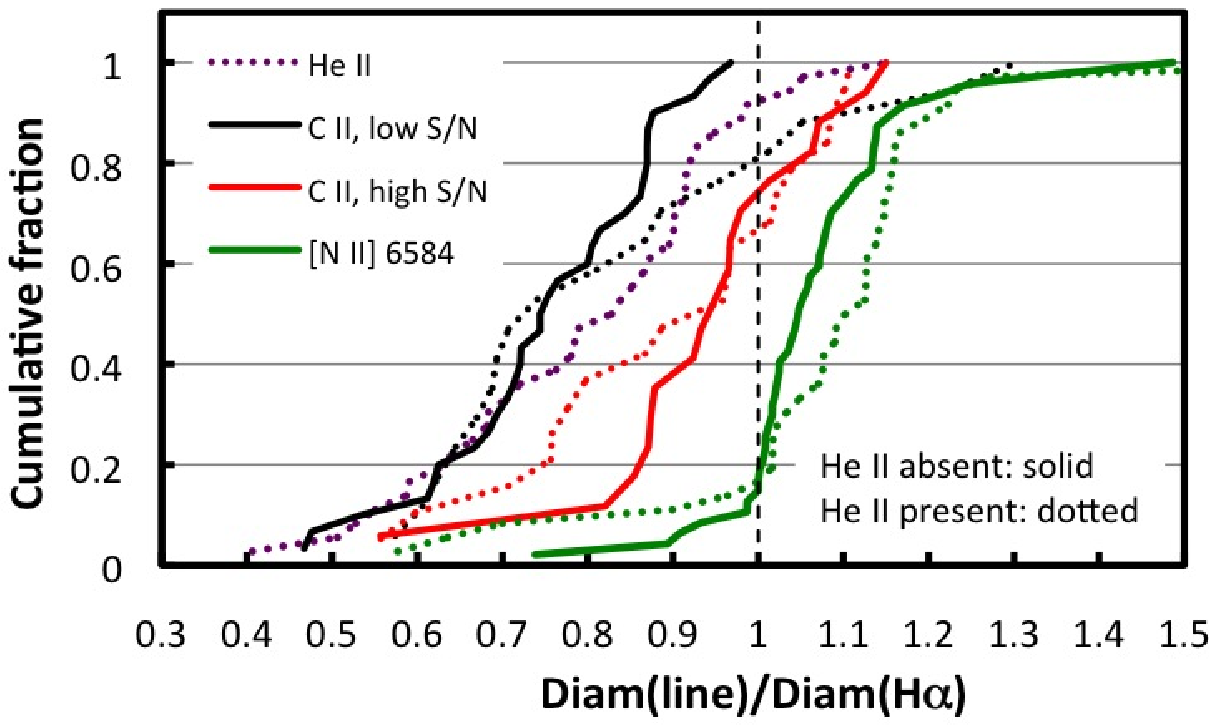}
\includegraphics[width=\columnwidth]{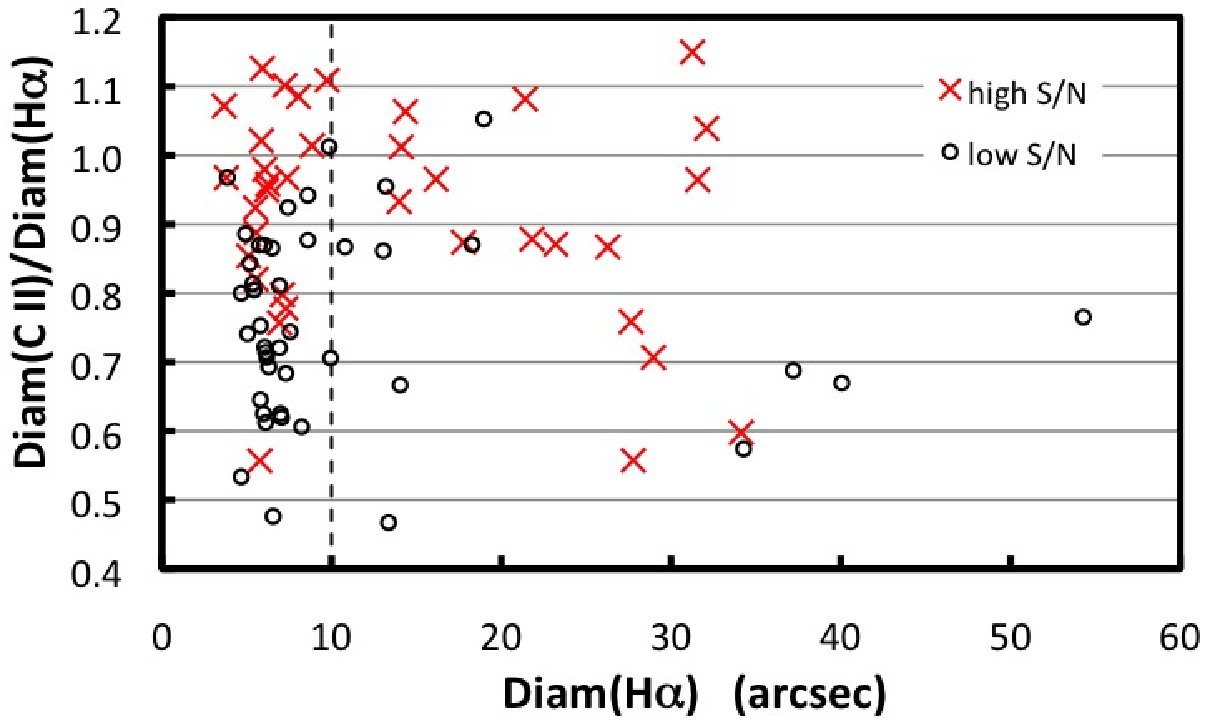}
\caption{Top:  We compare the cumulative distributions of the relative sizes of the planetary nebulae in all lines.  Clearly, the S/N affects the \ion{C}{2} $\lambda$6578 measurements.  The presence of \ion{He}{2} $\lambda$6560 has no effect on the relative size distributions.  The \ion{He}{2} $\lambda$6560 emission has the smallest spatial extent, [\ion{N}{2}] $\lambda$6583 the largest, while \ion{C}{2} $\lambda$6578 (high S/N) is intermediate.  Bottom:  The relative sizes measured in the \ion{C}{2} $\lambda$6578 line span a similar range in H$\alpha$ sizes for the objects whose \ion{C}{2} $\lambda$6578 spatial profiles have high and low S/N.  The offset between the two groups implies that the two methods to measure the size of the region emitting \ion{C}{2} $\lambda$6578 are not equivalent.}
\label{fig_diameters}
\end{figure}

\subsection{\ion{C}{2} $\lambda$6578 diameters}

The top panel in Figure \ref{fig_diameters} presents the spatial extents of the planetary nebulae from Table \ref{table_diameters} graphically.  Here, we plot the cumulative distribution of the ratio of the diameters in the \ion{He}{2} $\lambda$6560, \ion{C}{2} $\lambda$6578, and [\ion{N}{2}] $\lambda$6583 lines relative to the diameter measured in H$\alpha$ (henceforth, relative diameters).  We use relative diameters since they cancel the effect of the distances to the objects in our sample.
We see that the relative diameters are typically smallest for the \ion{He}{2} $\lambda$6560 line, which is almost always confined to a smaller spatial extent than H$\alpha$, i.e., approximately 90\% of the sample have a relative diameter less than 1.0.  The [\ion{N}{2}] $\lambda$6583 line presents the largest spatial extent, being almost always larger than H$\alpha$ (approximately 80\% have relative diameters exceeding 1.0).  For [\ion{N}{2}] $\lambda$6583, the presence or absence of the \ion{He}{2} $\lambda$6560 line has no effect upon the relative size compared to H$\alpha$.  

\begin{figure*}
\center{
\includegraphics[width=\columnwidth]{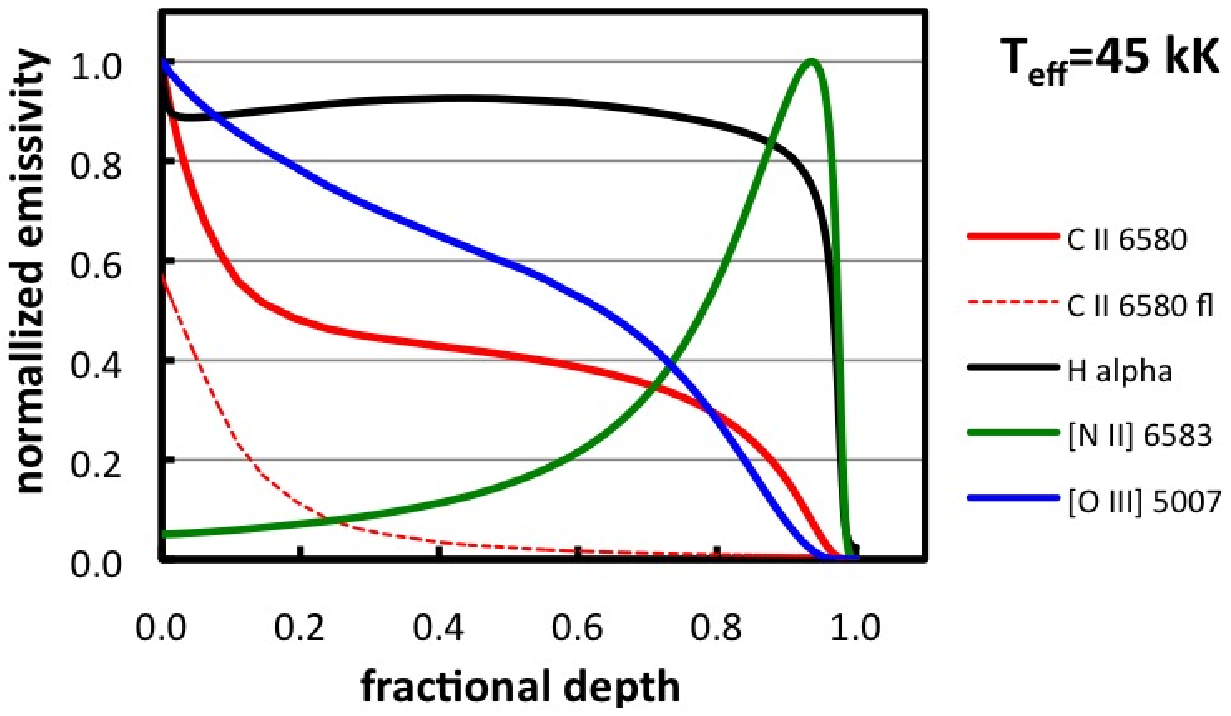} \includegraphics[width=\columnwidth]{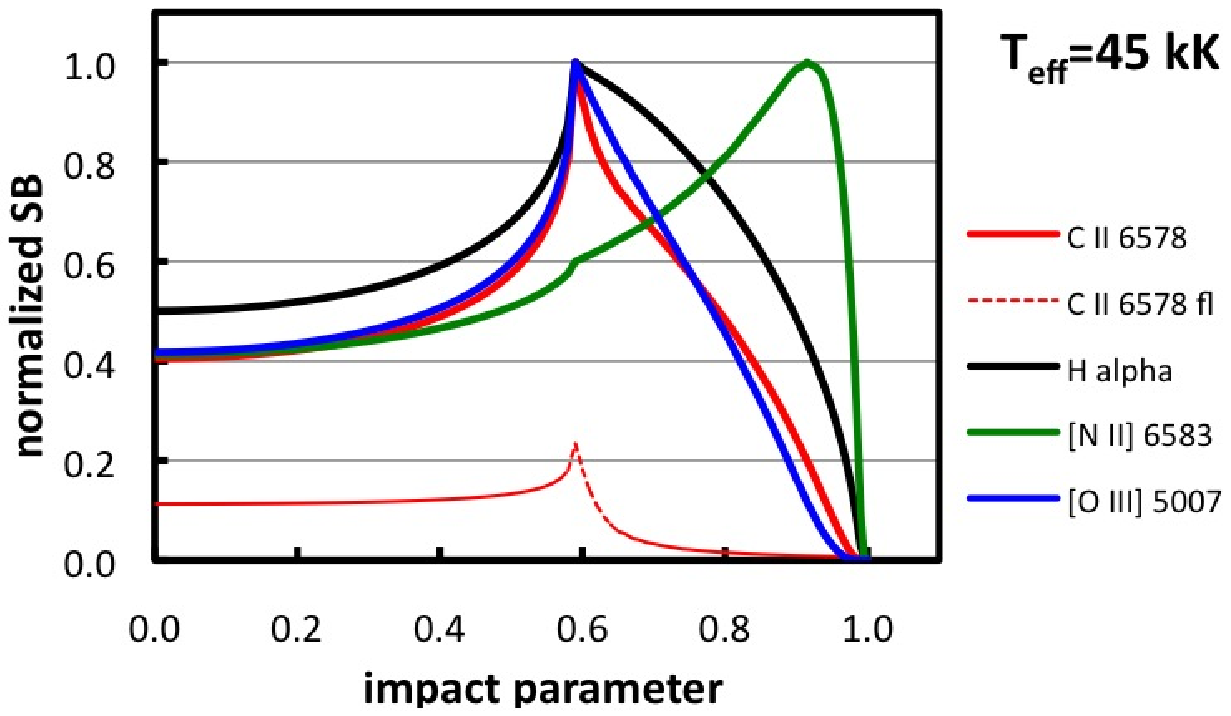}\\
\includegraphics[width=\columnwidth]{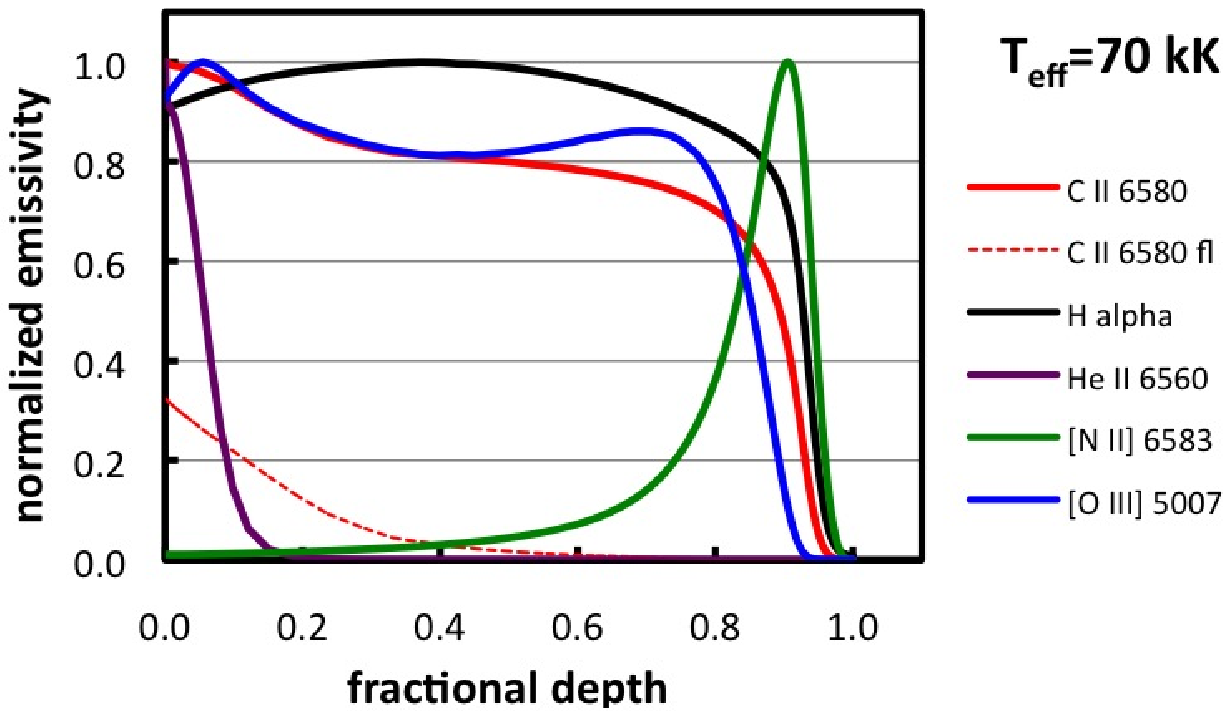} \includegraphics[width=\columnwidth]{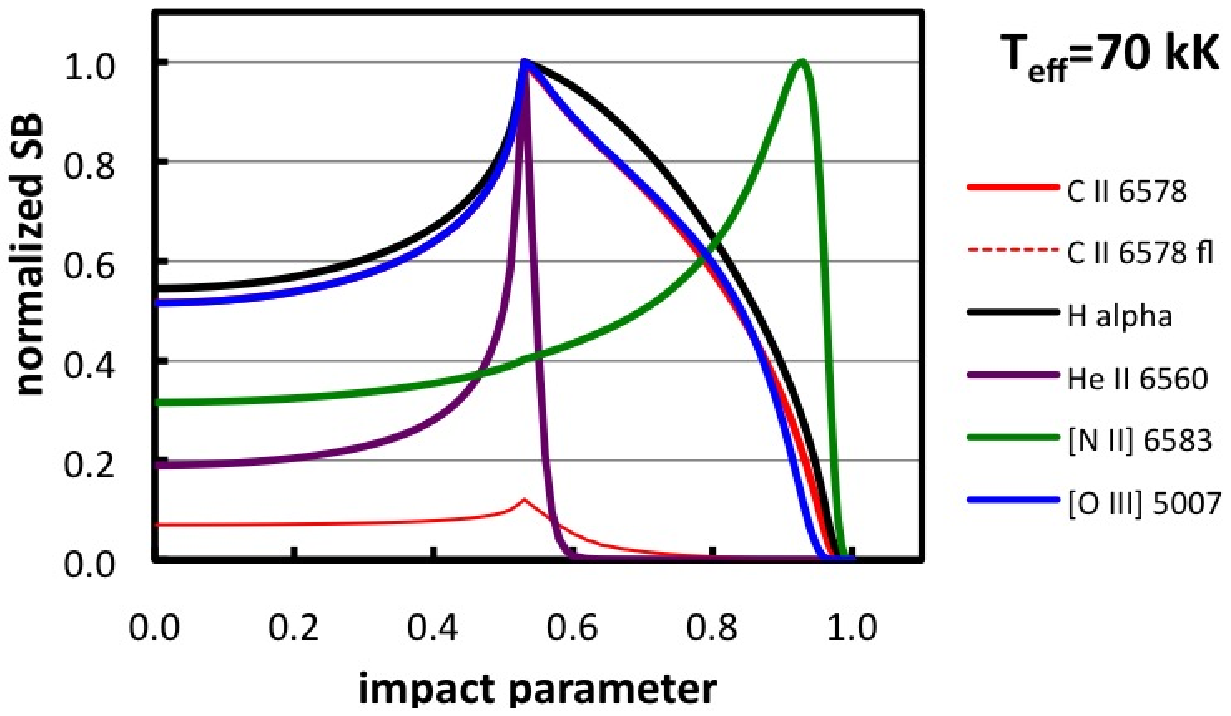}\\
\includegraphics[width=\columnwidth]{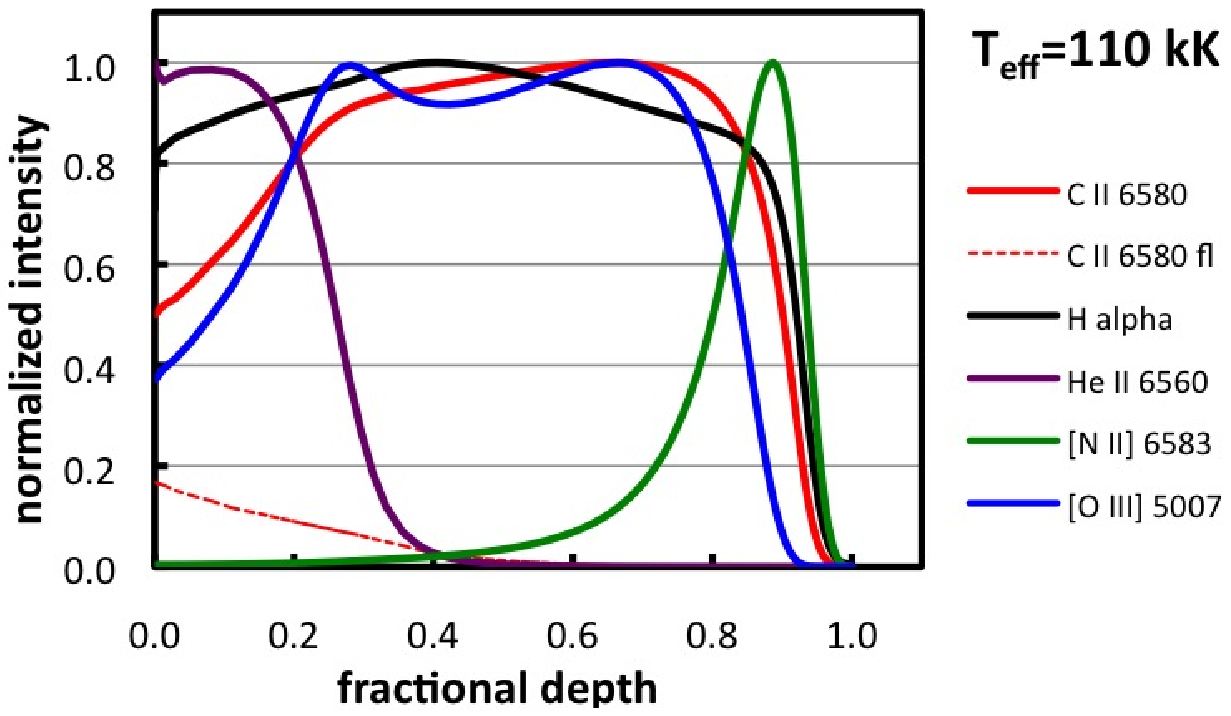} \includegraphics[width=\columnwidth]{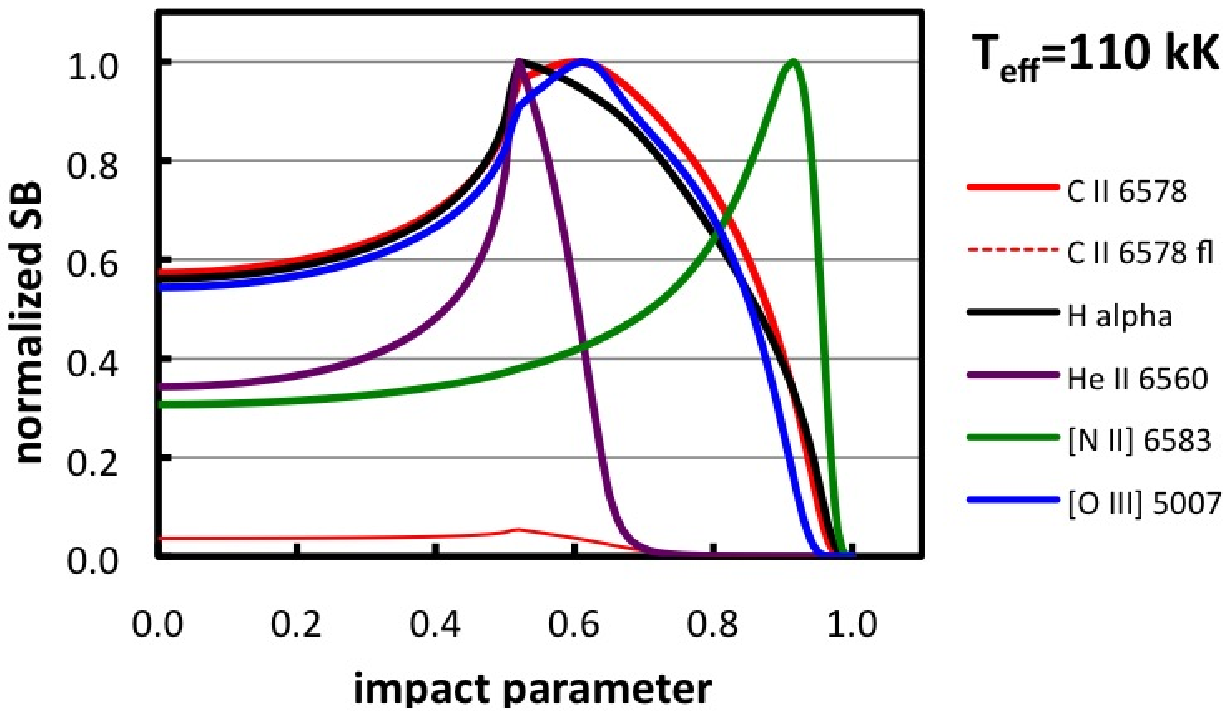} }
\caption{We present the normalized emissivity as a function of depth into the nebular shell (left column) and the normalized surface brightness as a function of impact parameter for three nebular models that differ only in the temperature of the central star (see panels) for the lines studied here: \ion{C}{2} $\lambda$6578, H$\alpha$, \ion{He}{2} $\lambda$6560, [\ion{N}{2}] $\lambda$6583, and [\ion{O}{3}] $\lambda$5007.  We computed these model nebulae using v13.08 of CLOUDY, last described by \citet{ferlandetal2013}, assuming a blackbody of luminosity $\log L_{bol} = 36.8$\,dex, an inner radius of $10^{17}$ cm, a constant density of 4000 cm$^{-3}$, a filling factor of unity, and CLOUDY's default abundance set for planetary nebulae.  For the \ion{C}{2} $\lambda$6578 line, we present the total emissivity/surface brightness (thick line) and the fraction of the emissivity/surface brightness due to fluorescence (thin line).  We do not show the \ion{He}{2} $\lambda$6560 line for the model with the coolest central star since it is generally not visible unless the central stars have a temperature of at least 70\,kK.}
\label{fig_models}
\end{figure*}

For the diameters measured in \ion{C}{2} $\lambda$6578, we must consider the S/N of the spatial profile.  There is a clear, statistically significant difference between the distributions for the objects with high and low S/N (Figure \ref{fig_diameters}; top panel).  If we compare the distributions of relative diameters for the objects without \ion{He}{2} $\lambda$6560 emission and test the hypothesis that the high S/N sample has larger diameters, a U test supports it, indicating a probability of only $1.4\times 10^{-5}$ of obtaining the two samples randomly from the same parent population.  In the bottom panel in Figure \ref{fig_diameters}, we plot the relative diameter in \ion{C}{2} $\lambda$6578 as a function of the diameter measured in H$\alpha$.  There is a clear offset between the relative diameters in \ion{C}{2} $\lambda$6578 for the two measurement methods, even though they span the same range of diameters in H$\alpha$, so we cannot consider the two measurement methods equivalent.  In what follows, we do not use the subset of diameters measured in \ion{C}{2} $\lambda$6578 with low S/N, except to make an internal test of our data and for which its low S/N is an asset (\S 4.5, Figure \ref{fig_high_s2n_c2}).

Considering only the high S/N sample in the top panel of Figure \ref{fig_diameters}, we see that the \ion{C}{2} $\lambda$6578 emission spans a spatial extent intermediate between those of \ion{He}{2} 6560 and [\ion{N}{2}] $\lambda$6583.  Usually, the \ion{C}{2} 6578 emission spans a smaller spatial extent than that of H$\alpha$, since 75\% of the sample has a relative diameter less than unity, and, as for [\ion{N}{2}] $\lambda$6583, the relative diameter of the \ion{C}{2} 6578 emission does not depend strongly upon the absence or presence of \ion{He}{2} $\lambda$6560.  

\section{Analysis}

\subsection{Nebular models}
\label{section_models}

\begin{figure}[t]
\includegraphics[width=\columnwidth]{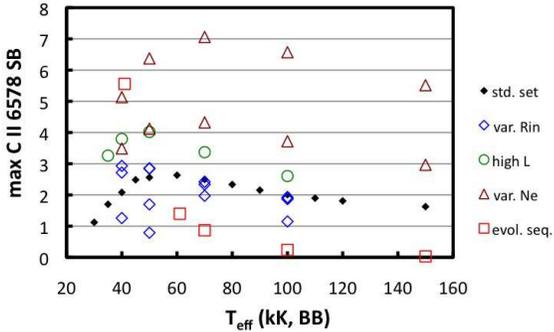}
\caption{We present the maximum surface brightness (arbitrary units) in the \ion{C}{2} $\lambda$6578 line for the full grid of models.  The models presented in Figure \ref{fig_models} belong to the standard set.  The other model sequences differ with respect to the standard set by varying the inner radius (0.25, 0.5, 2.0, and 5.0) in units of $10^{17}$ cm, higher luminosity, $\log L_{bol} = 37.3$\,dex, higher density, 6,300 cm$^{-3}$ (lower sequence) and 10,000 cm$^{-3}$ (higher sequence), and a sequence that approximates (with constant density) the solar metallicity model in Figures 5-7 from \citet[][the models at 60\,kK and 70\,kK are interpolations]{schonberneretal10}.}
\label{fig_grid_models}
\end{figure}

To provide context for our results and a framework for the analysis that follows, we consider a grid of generic model nebulae, of which we present three in Figure \ref{fig_models}.  These models are meant to explore the range of properties we expect for objects in our sample (details in the figure caption), not to represent individual objects.  In the left column, the normalized emissivity is plotted as a function of fractional depth into the nebular shell for the lines of \ion{C}{2} $\lambda$6578, H$\alpha$, \ion{He}{2} $\lambda$6560, [\ion{N}{2}] $\lambda$6583, and [\ion{O}{3}] $\lambda$5007.  On the right, the normalized surface brightness is plotted as a function of impact parameter for the same lines.  For the \ion{C}{2} $\lambda$6578 line, both the total emissivity/surface brightness and the fraction contributed by fluorescence are shown.  Figure \ref{fig_grid_models} presents the maximum surface brightness in the \ion{C}{2} $\lambda$6578 line for the entire grid (additional model details in the figure caption).  

From Figure \ref{fig_models}, the emissivity of the \ion{C}{2} $\lambda$6578 line follows that for [\ion{O}{3}] $\lambda$5007 closely, which is also true for the full grid of models.  Both the  \ion{C}{2} $\lambda$6578 and [\ion{O}{3}] $\lambda$5007 lines span the extent of the zone emitting in H$\alpha$, and sample it especially well for the models whose central stars have higher temperatures (e.g., middle row).  Figure \ref{fig_grid_models} shows that it is possible to choose model parameters that produce a high surface brightness in the \ion{C}{2} $\lambda$6578 line for central stars spanning at least the 40\,kK--150\,kK temperature range.  However, high \ion{C}{2} $\lambda$6578 surface brightnesses are favored for high density, high luminosity, and, for low central star temperatures, compact nebulae.  Of these three parameters, density has the strongest effect and nebular size the weakest.  Fluorescence may contribute a significant fraction of the total emissivity in the \ion{C}{2} $\lambda$6578 line, especially for models with low temperature central stars (Figure \ref{fig_models}, top row), but, even in these cases, the contribution from fluorescence never dominates the total surface brightness.  

Although Figure \ref{fig_models} provides no kinematic information, the general result is well-known \citep[e.g.,][]{wilson1950}.  The nebular plasma behaves as a fluid, so the inner parts may not overrun the outer parts \citep[e.g.,][]{villaveretal2002,perinottoetal2004}.  We expect lower velocities and so smaller line widths from the innermost regions, e.g., for the \ion{He}{2} $\lambda$6560 line, and larger line widths from regions farther out, e.g., for the [\ion{N}{2}] $\lambda\lambda$6548,6583 lines at the outer edge of the ionized shell.  Therefore, these models would predict similar line widths for the \ion{C}{2} $\lambda$6578, [\ion{O}{3}] $\lambda$5007, and H$\alpha$ lines, given that they sample similar volumes of the nebular shells.  The \ion{He}{2} $\lambda$6560 line should be narrower, for it samples only the innermost fraction of the nebular volume.  Finally, the [\ion{N}{2}] $\lambda\lambda$6548,6583 lines should be widest because they arise primarily from the outermost part of the nebular volume.

\begin{figure*}
\center{
\includegraphics[width=\columnwidth]{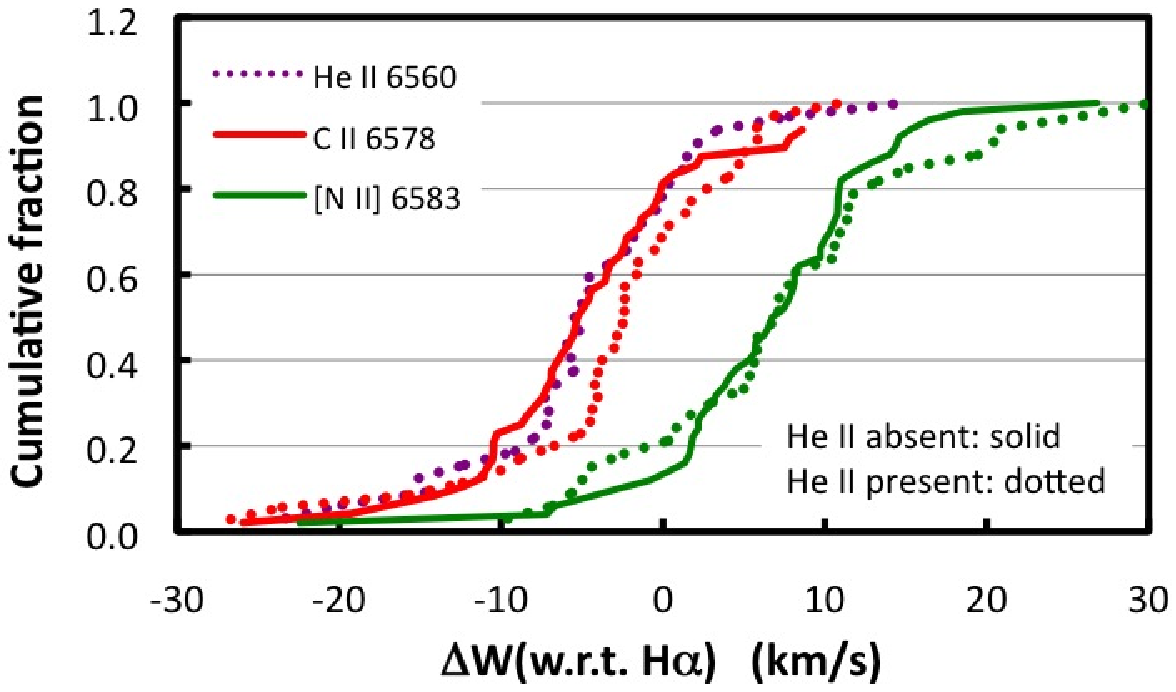} \includegraphics[width=\columnwidth]{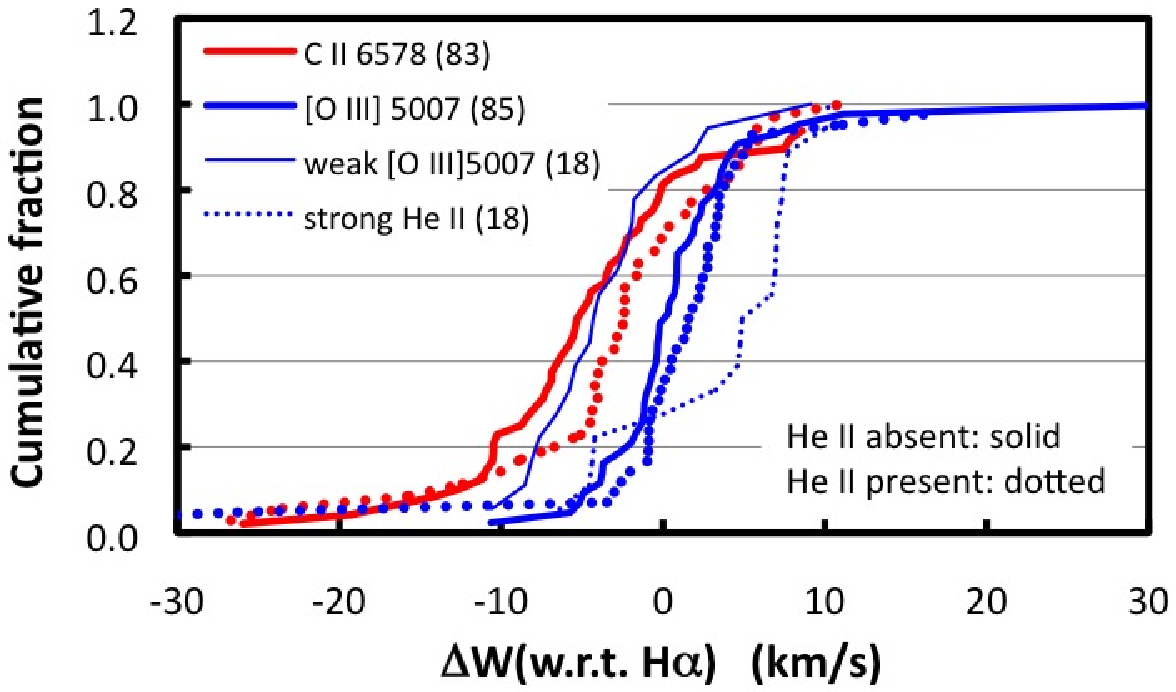}
}
\caption{left panel:  We present the cumulative distributions of the relative line widths (with respect to H$\alpha$, see text) of \ion{He}{2} $\lambda$6560, \ion{C}{2} $\lambda$6578, and [\ion{N}{2}] $\lambda$6583.  Note that the relative line widths of \ion{C}{2} $\lambda$6578 and [\ion{N}{2}] $\lambda$6583 do not depend upon whether \ion{He}{2} $\lambda$6560 is absent (solid lines) or present (dotted lines).  right panel:  We present the cumulative distributions of the relative line widths for \ion{C}{2} $\lambda$6578 (this study) and [\ion{O}{3}] $\lambda$5007 \citep[][]{richeretal2008,richeretal2010}.  The two samples plotted with thick lines from \citet[][]{richeretal2008,richeretal2010} are in same evolutionary phases as the objects in the present sample while the \lq\lq weak [\ion{O}{3}]" and \lq\lq strong \ion{He}{2}" samples are in earlier (low ionization) and later (very high ionization) phases, respectively, and illustrate how different evolutionary phases sample the zone emitting in H$\alpha$ in different ways.  The numbers in brackets in the legend indicate the number of objects in each sample of planetary nebulae.  
}
\label{fig_dfwhm_all}
\end{figure*}

\subsection{The \ion{C}{2} $\lambda$6578 kinematics}
\label{sec_c2kin}

As a first approximation to investigate whether the kinematics of the \ion{C}{2} $\lambda$6578 line are those expected, we compare the cumulative distribution of its line width with the cumulative distribution of line widths for H$\alpha$, both of which are shown in Figure \ref{fig_fwhm_evol}.  Our justification for doing so is that the models in Figure \ref{fig_models} indicates that \ion{C}{2} $\lambda$6578 emission arises throughout the zone from which H$\alpha$ emission arises.  We then test the null hypothesis that the cumulative distribution of H$\alpha$ line widths is shifted to higher values with respect to that of the \ion{C}{2} $\lambda$6578 line widths.  Applying the U test supports this hypothesis, yielding a probability of only 0.75\% that the two distributions arise by chance from the same parent population.  From this result, we conclude that the line widths of the \ion{C}{2} $\lambda$6578 line are not those expected from ionization equilibrium in a chemically-homogeneous plasma.  Since the \ion{C}{2} $\lambda$6578 line widths are narrower than expected, the kinematics of the \ion{C}{2} $\lambda$6578 line correspond to a zone more internal in the nebula than expected.  

However, it is feasible to devise more sensitive tests of whether the kinematics of the \ion{C}{2} $\lambda$6578 line correspond to expectations.  Figures \ref{fig_fwhm_correlations} (top and bottom panels) and \ref{fig_fwhm_evol} emphasize that an analysis of the kinematics of this sample must account for the changes in the kinematics as the planetary nebulae evolve.  We do so by comparing the difference in line widths with respect to the H$\alpha$ line, i.e., $\mathrm{FWHM}(\mathrm{line}) - \mathrm{FWHM}(\mathrm H\alpha)$.  Henceforth, we refer to these differences in line widths as relative line widths.  Since H$\alpha$ arises from the entire volume in all objects, the relative line widths allow us to compare the kinematics of the sub-volume that gives rise to a given emission line with respect to the kinematics of the entire ionized shell.  

We present the relative line widths for the \ion{He}{2} $\lambda$6560, \ion{C}{2} $\lambda$6578, and [\ion{N}{2}] $\lambda$6583 lines in Figure \ref{fig_dfwhm_all} (left panel).  This figure summarizes the three panels in Figure \ref{fig_fwhm_correlations} and reflects the discussion of the kinematics of the nebular models in the previous section.  As previously found, the distribution of relative line widths for the \ion{He}{2} $\lambda$6560 line is narrowest, that for the [\ion{N}{2}] $\lambda$6583 line is the widest, and the distribution for the \ion{C}{2} $\lambda$6578 line is intermediate, though just barely.  Note that the evolutionary effect observed in Figure \ref{fig_fwhm_evol} is eliminated for the \ion{C}{2} $\lambda$6578 and [\ion{N}{2}] $\lambda$6583 lines.  We also note that the distributions of the relative line widths for the \ion{He}{2} $\lambda$6560 and \ion{C}{2} $\lambda$6578 lines are indistinguishable.  This is not expected from the models (\S\ref{section_models}):  Since the \ion{C}{2} $\lambda$6578 line samples the kinematics of a much larger fraction of the nebular shell than the \ion{He}{2} $\lambda$6560 line, the relative line width of \ion{C}{2} $\lambda$6578 would be expected to be larger.   

In addition, the models in Figure \ref{fig_models} clearly indicate that the way the \ion{C}{2} $\lambda$6578 emission probes the volume from which it arises depends upon the temperature of the central star.  For cool central stars, the \ion{C}{2} $\lambda$6578 emission comes primarily from the inner part of the zone that emits this line (Figure \ref{fig_models}, top row).  As the central star temperature increases, the \ion{C}{2} $\lambda$6578 emission probes the volume from which it arises more uniformly (Figure \ref{fig_models}, middle row).  At the hottest temperatures, the \ion{C}{2} $\lambda$6578 emission is no longer a very good probe of the central part of the volume from which it arises (Figure \ref{fig_models}, bottom row).  This is observed in the right panel of Figure \ref{fig_dfwhm_all}, where we use the relative line widths [\ion{O}{3}] $\lambda$5007 \citep{richeretal2008,richeretal2010} to illustrate the effect.  These four samples of planetary nebulae from \citet{richeretal2008,richeretal2010} were chosen to select four evolutionary phases up to the cessation of nuclear burning in their central stars (maximum temperature).  The heavy lines (both solid and dotted) present the planetary nebula in evolutionary phases most similar to those observed here for \ion{C}{2} $\lambda$6578 while the thin lines represent samples of planetary nebulae deliberately chosen to be very young/unevolved (\lq\lq weak [\ion{O}{3}]") and more highly evolved (\lq\lq strong \ion{He}{2}").  

The right panel in Figure \ref{fig_dfwhm_all} indicates that the cumulative distribution of the relative line widths of [\ion{O}{3}] $\lambda$5007 do depend upon the evolutionary phase.  The youngest, least evolved planetary nebulae with the coolest central stars (below 40-45\,kK based upon the models considered here) have differential line widths that are significantly negative, as expected from Figure \ref{fig_models} (top), since the [\ion{O}{3}] $\lambda$5007 emission arises primarily in the inner part of the nebular shell.  Conversely, the most evolved planetary nebulae containing the hottest stars (probably exceeding 120\,kK and being optically thin based upon the models considered here) have differential line widths that are significantly positive.  This happens, not because the [\ion{O}{3}] $\lambda$5007 emission arises external to the volume that emits H$\alpha$, but instead because it does not sample well the innermost part of the volume giving rise to H$\alpha$.  Between these extremes, the distribution of relative line widths is very similar to that of H$\alpha$, independently of whether \ion{He}{2} emission is absent or present (\ion{He}{2} emission arises for central stars with temperatures above approximately 70\,kK for the models considered here).  

\begin{figure}[t]
\includegraphics[width=\columnwidth]{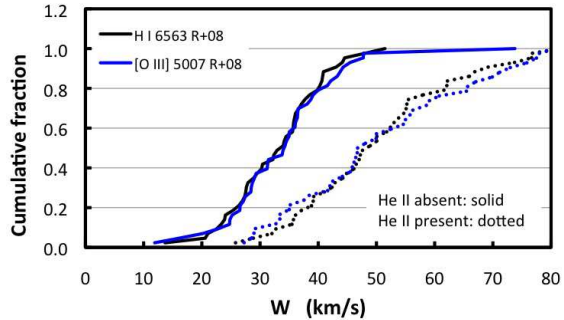}
\caption{We compare the cumulative distributions of line widths for the H$\alpha$ and [\ion{O}{3}] 5007 lines for the planetary nebulae from \citet{richeretal2008}, segregating the objects according to the absence or presence of the \ion{He}{2} $\lambda$6560 line.  The cumulative distributions for the two lines are statistically indistinguishable, regardless of the presence or absence of the \ion{He}{2} $\lambda$6560 line.}
\label{fig_R08data}
\end{figure}

Both panels of Figure \ref{fig_dfwhm_all} indicate that the kinematics of the \ion{C}{2} $\lambda$6578 line are not those expected.  The kinematics of the \ion{C}{2} $\lambda$6578 line mimic the kinematics expected of emission from the innermost part of the nebular shell, probed either by the \ion{He}{2} $\lambda$6560 line in objects with hotter central stars ($> 70$\,kK) or the [\ion{O}{3}] $\lambda$5007 line in planetary nebulae with the coolest central stars.  If we adopt the null hypothesis that the distribution of relative line widths in [\ion{O}{3}] $\lambda$5007 is shifted to higher values than that for the \ion{C}{2} $\lambda$6578 line, a U-test supports it, finding a minuscule probability ($4.2\times 10^{-7}$) that the two distributions arise by chance from the same parent distribution.  Of the planetary nebulae in our sample, 23 also belong to the samples of \citet{richeretal2008,richeretal2010}.  We may thus inquire whether the distributions of relative [\ion{O}{3}] $\lambda$5007 line widths differ between the objects with \ion{C}{2} $\lambda$6578 emission and those without it among the objects from \citet{richeretal2008,richeretal2010}.  A U-test indicates that there is a probability of 11.4\% of obtaining the two distributions by chance from the same parent distribution, and so they are not statistically different.  Thus, to the extent that the distribution of relative line widths in [\ion{O}{3}] $\lambda$5007 from \citet{richeretal2008, richeretal2010} indicate the expected distribution of relative line widths for the \ion{C}{2} $\lambda$6578 line, the kinematics we observe for \ion{C}{2} $\lambda$6578 do not match expectations based upon ionization equilibrium in chemically-homogeneous nebulae.

To support our approximation that the kinematics of the H$\alpha$ line should be very similar to those of the [\ion{O}{3}] $\lambda$5007 line (and the \ion{C}{2} $\lambda$6578 line by extension), Figure \ref{fig_R08data} compares the cumulative distributions of the line widths for H$\alpha$ and [\ion{O}{3}] $\lambda$5007, based upon the data from \citet[][]{richeretal2008}.  Their study employed an analysis similar to that used here.  However, in order to observe H$\alpha$ and [\ion{O}{3}] $\lambda$5007, \citet{richeretal2008} had to acquire separate spectra of each wavelength region for each object.  Although these H$\alpha$ and [\ion{O}{3}] $\lambda$5007 spectra were almost always obtained consecutively, it is possible that they suffer from small spatial offsets and so the two spectra might sample slightly different volumes within each object.  Despite these caveats, which do not apply to our \ion{C}{2} $\lambda$6578 sample, the distributions of line widths for H$\alpha$ and [\ion{O}{3}] $\lambda$5007 are statistically indistinguishable, both when \ion{He}{2} $\lambda$6560 line is absent and present.  Thus, the previous finding that the kinematics of the zones emitting \ion{C}{2} $\lambda$6578 and H$\alpha$ differ should be robust.  

We may also investigate whether the distribution of relative line widths for \ion{C}{2} $\lambda$6578 behave as expected by comparing them directly with what would be expected if they traced the kinematics of the H$\alpha$ line perfectly (though including measurement uncertainties).  If that were the case, we would expect that the median value of the differential line widths would be $0.0$\,km/s, but instead we find $-3.6$\,km/s.  Indeed, 76\% of the \ion{C}{2} $\lambda$6578 differential line widths are negative, not 50\% as would be expected if the kinematics of the H$\alpha$ and \ion{C}{2} $\lambda$6578 lines were perfectly correlated.  We can test whether this difference is statistically significant by computing the $\chi^2$ statistic after binning the objects according to whether their \ion{C}{2} $\lambda$6578 relative line width is negative or positive and comparing with an expected evenly split distribution.  The $\chi^2$ statistic so obtained is 22.3.  The probability of this result is miniscule, $2.4\times 10^{-6}$, so we may once more conclude that the kinematics of the \ion{C}{2} $\lambda$6578 line are not those expected based upon ionization equilibrium.  (Applying the same test to the [\ion{O}{3}] $\lambda$5007 differential line widths for the planetary nebulae in similar evolutionary phases in Figure \ref{fig_dfwhm_all} returns the result that the distributions of differential line widths in H$\alpha$ and [\ion{O}{3}] $\lambda$5007 are not statistically distinct.)

Thus, it appears that the kinematics of the \ion{C}{2} $\lambda$6578 line are not consistent with expectations based upon ionization equilibrium (Figure \ref{fig_models}).  In particular, the plasma giving rise to the \ion{C}{2} $\lambda$6578 line has the kinematics expected for a nebular volume that does not sample the full volume from which H$\alpha$ arises, but rather the innermost part of it.  Therefore, the kinematics of the \ion{C}{2} $\lambda$6578 line are like those of a line of a higher ionization stage.  This need not imply that this plasma is co-spatial with higher ionization stages, but only that its kinematics mimic the kinematics of a plasma of a higher ionization stage.   

In \S\ref{sec_introduction} we noted that one of the better-explored explanations for the abundance discrepancy problem was the existence of cold, hydrogen-poor plasma intermixed with the normal nebular plasma \citep[e.g.,][]{liuetal2000}, yet we assumed a temperature of $10^4$\,K to correct for the thermal broadening of the \ion{C}{2} $\lambda$6578 line in \S\ref{sec_methodology}.  If the \ion{C}{2} $\lambda$6578 line arose from cold plasma, this correction for thermal broadening would be inappropriate.  To check that all is well with our statistical analyses, we recalculated all of these results adopting no correction for thermal broadening in the \ion{C}{2} $\lambda$6578 line, and all of the results remain unchanged, compliant with the criterion declared in \S\ref{sec_c2_line_widths}.  The test that is most affected is the hypothesis that the distribution of H$\alpha$ line widths is shifted to higher values than that for \ion{C}{2} $\lambda$6578.  The U-test still supports the hypothesis, but the probability that two distributions may arise by chance from a single progenitor population rises to 1.7\% (from 0.75\%).  Thus, all of the results presented above are independent of the correction for thermal broadening for the \ion{C}{2} $\lambda$6578 line.

\subsection{The size of the \ion{C}{2} $\lambda$6578 emission region}

Unfortunately, we are unable to come to any conclusion regarding the spatial extent of the \ion{C}{2} $\lambda$6578 emission region.  Our grid of models indicates that diameter of the \ion{C}{2} $\lambda$6578 emission region should be \emph{smaller} than that emitting in H$\alpha$ by $1-3$\%.  For the [\ion{N}{2}] $\lambda$6583 emission region, the models indicate diameters that are $1-2$\% \emph{larger} than the region emitting in H$\alpha$.  The observed median values are 5\% smaller and 7\% larger for the diameters in \ion{C}{2} $\lambda$6578 and [\ion{N}{2}] $\lambda$6583, respectively.  Both differences would be highly significant, but are likely spurious since the constant density structure in our grid of models is not likely to be a good approximation to reality.   

This limitation will not affect the kinematics discussed in the previous section since that depends essentially on the physics involving the ions of H$^+$, C$^{++}$, and O$^{++}$ (ionization, recombination, fluorescence, and collisional excitation).

\subsection{When is \ion{C}{2} $\lambda$6578 emission present?}
\label{section_cii_presence}

\begin{figure}[t]
\includegraphics[width=\columnwidth]{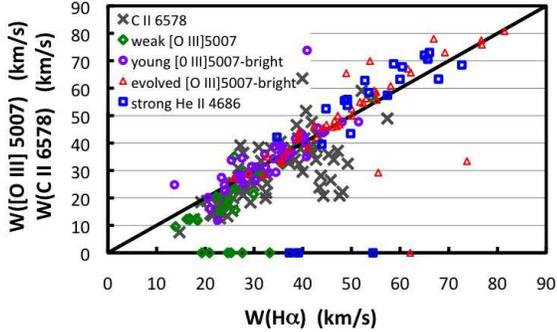}
\caption{We superpose the correlation of the line widths of [\ion{O}{3}] $\lambda$5007 and H$\alpha$ for the planetary nebulae from \citet[][]{richeretal2008,richeretal2010} on the correlation of the line widths of \ion{C}{2} $\lambda$6578 and H$\alpha$ presented in the top panel of Figure \ref{fig_fwhm_correlations}.  The planetary nebulae with \ion{C}{2} $\lambda$6578 emission have H$\alpha$ line widths that span a considerably smaller range than those from \citet{richeretal2008,richeretal2010}.  }
\label{fig_fwhm_o3_c2}
\end{figure}

\begin{figure}[t]
\includegraphics[width=\columnwidth]{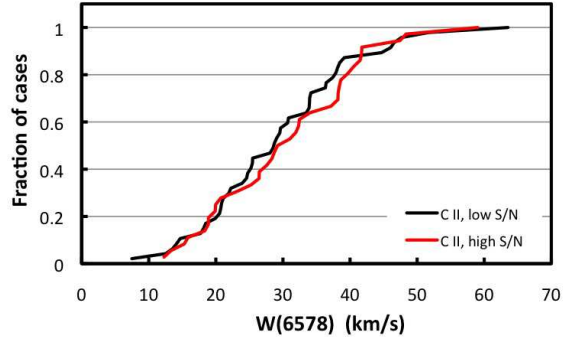}
\caption{We compare the distribution of line widths for the objects whose \ion{C}{2} $\lambda$6578 spatial profiles have low and high S/N.  If the SPM Catalogue lacked the sensitivity to detect wider \ion{C}{2} $\lambda$6578 lines than observed, we would expect a lack of the largest line widths for spatial profiles with low S/N, but we find no difference between the two distributions.}
\label{fig_high_s2n_c2}
\end{figure}

\begin{figure}[t]
\includegraphics[width=\columnwidth]{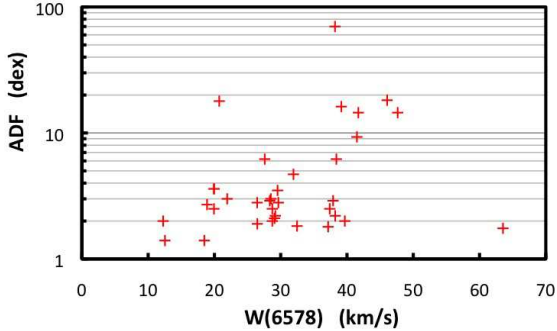}
\caption{We plot the ADFs for oxygen from Table \ref{table_diameters} as a function of the \ion{C}{2} 6578 line width from Table \ref{table_velocities}.  There is no clear trend.}
\label{fig_adf_c2lw}
\end{figure}

\begin{figure*}
\begin{center}
\includegraphics[width=2.0\columnwidth]{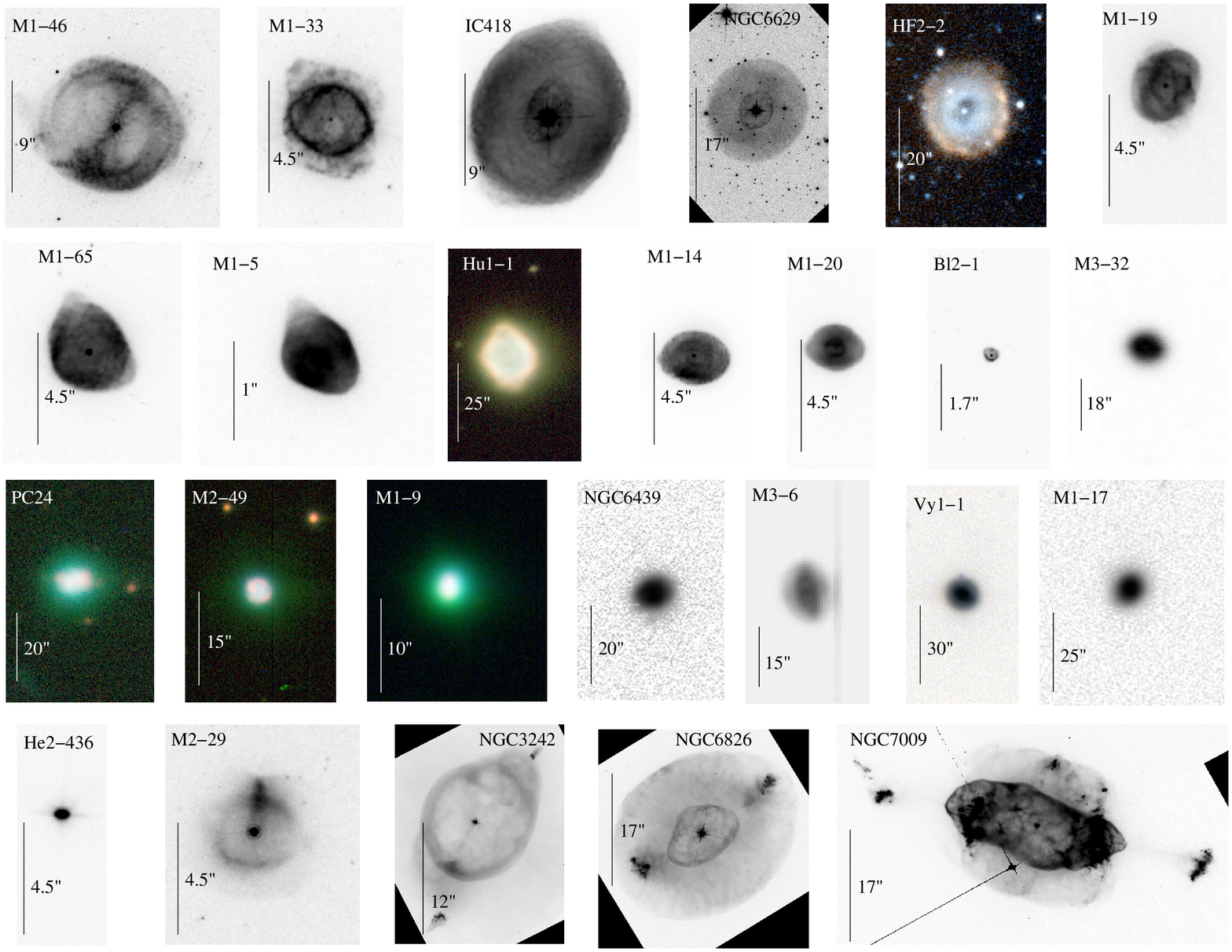}  
\includegraphics[width=2.0\columnwidth]{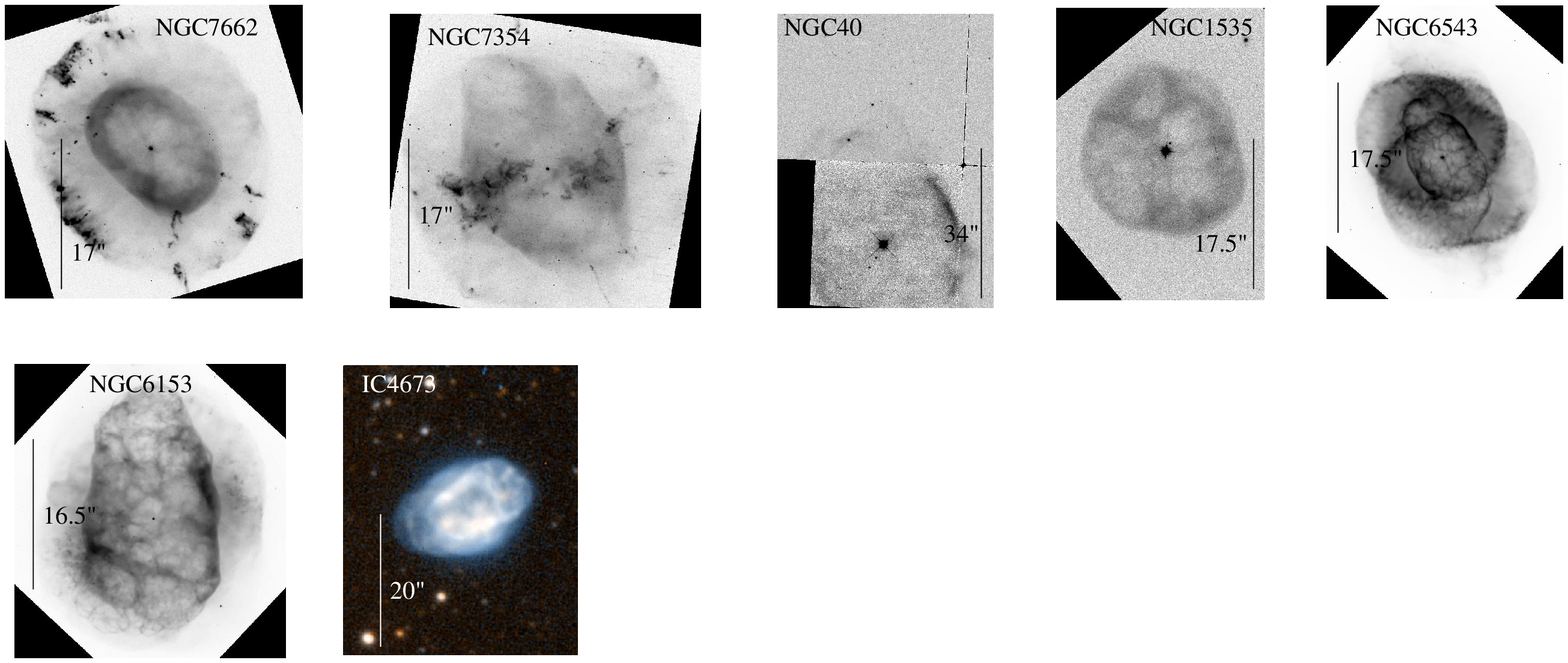}  
\end{center}
\caption{We present a montage of the planetary nebulae in our sample that have elliptical-like morphologies.  North is up and East to the left.  The bar in each panel indicates the spatial scale.  Most of these images are from the Hubble Legacy Archive, but those of Hf 2-2, Hu 1-1, IC 4673, M 1-8, M 1-17, M 2-19, M 2-49, M 3-32, NGC 6778, PC 24, Vy 1-1 are from the IAC Morphological Catalog of Northern Galactic Planetary Nebulae \citep[][]{manchadoetal1996}, while that of NGC 6439 comes from the SDSS DR12 \citep{alametal2015} and that of M 3-6 is from our imaging \citep{lopezetal2012}.  }
\label{fig_mosaic_elliptical}
\end{figure*}

\begin{figure*}
\begin{center}
\includegraphics[width=2.0\columnwidth]{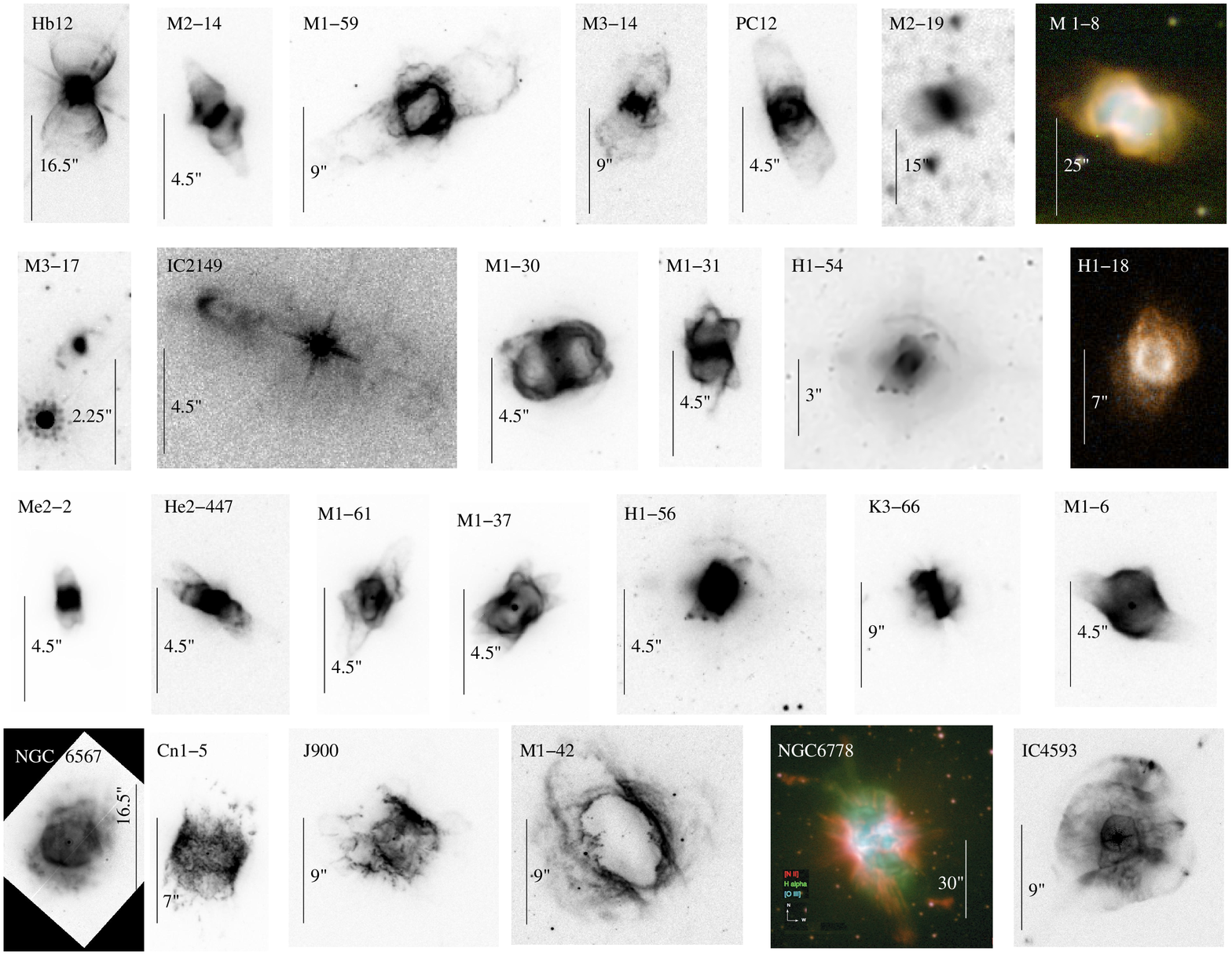}  
\end{center}
\caption{We present a montage of the planetary nebulae in our sample that have bipolar or multipolar morphologies.  North is up and East to the left.  The bar in each panel indicates the spatial scale.  Fig. \ref{fig_mosaic_elliptical} provides references for the images presented here.}
\label{fig_mosaic_bipolar}
\end{figure*}

In Figure \ref{fig_fwhm_o3_c2} we superpose on the top panel of Figure \ref{fig_fwhm_correlations} the correlation of the line widths of [\ion{O}{3}] $\lambda$5007 and H$\alpha$ lines for the planetary nebulae in the Milky Way bulge from \citet[][]{richeretal2008,richeretal2010}.  Our intention is to attempt to understand whether the \ion{C}{2} $\lambda$6578 line arises during all phases of evolution of planetary nebulae.  The planetary nebulae from \citet[][]{richeretal2008,richeretal2010} were selected to span all evolutionary phases from the onset of nebular ionization up to and slightly beyond the cessation of nuclear burning in the central stars.  Clearly, the H$\alpha$ line widths from planetary nebulae that present \ion{C}{2} $\lambda$6578 emission span a smaller range than those of the planetary nebulae from \citet[][]{richeretal2008,richeretal2010}.  In particular, there is a complete lack of planetary nebulae that present \ion{C}{2} $\lambda$6578 emission with the largest H$\alpha$ line widths observed by \citet[][]{richeretal2008,richeretal2010}.  

We note that \citet{richeretal2008,richeretal2010} measure line widths by fitting a single Gaussian profile to all of their objects, which is also the method used here for all but 14 spectra.  Had we used the corrected line width from the single Gaussian fit for these 14 spectra, the previous result would be unchanged.  In addition, \citet{richeretal2008,richeretal2010} always considered the full spatial extent of their spectra when making their measurements, thereby including more material with velocities near the systemic velocity, which should systematically bias their measurements to lower values than those that would be measured with the method used here (see \S \ref{sec_methodology}).  Thus, the difference noted above is all the more striking.  

Although we expect that the SPM Catalogue is limited due to surface brightness, it is unclear why the sample of planetary nebulae presenting \ion{C}{2} $\lambda$6578 emission lacks objects with large H$\alpha$ line widths.   In Figure \ref{fig_high_s2n_c2}, we compare the distribution of the \ion{C}{2} $\lambda$6578 line widths for the planetary nebulae whose \ion{C}{2} $\lambda$6578 spatial profiles have high and low S/N.  If a very large line width made detection of the \ion{C}{2} $\lambda$6578 line difficult, we would expect to see a deficit at the largest line widths for objects whose \ion{C}{2} $\lambda$6578 spatial profiles have low S/N, but there is no statistically significant difference between the distributions for these two samples.  Thus, it is unlikely that the data from the SPM Catalogue lacks the sensitivity to detect very wide \ion{C}{2} $\lambda$6578 lines due to the faintness of this line. 

From Figure \ref{fig_fwhm_o3_c2}, we see that, when plotted as a function of the H$\alpha$ line width, the dispersion of the \ion{C}{2} $\lambda$6578 line widths is greater than the dispersion in [\ion{O}{3}] $\lambda$5007 line widths.  Again, this indicates that, in general, there is less correlation between the volumes from which the H$\alpha$ and the \ion{C}{2} $\lambda$6578 lines are emitted than between the volumes from which the H$\alpha$ and [\ion{O}{3}] $\lambda$5007 lines are emitted, supporting the conclusions of \S\ref{sec_c2kin}.    

\citet{robertsontessigarnett2005} find a marginal correlation between expansion velocity and the ADF for the 22 planetary nebulae they studied.  We approximate the ADF as the ratio of total O or O$^{2+}$ ionic abundances calculated from permitted and collisionally-excited lines from literature sources (see Table \ref{table_diameters}).  Figure \ref{fig_adf_c2lw} presents the ADF from the literature plotted as a function of the line width measured for the \ion{C}{2} $\lambda$6578 line from Table \ref{table_velocities} (35 objects).  We find no clear correlation (and it does not improve if the H$\alpha$ line width is used).  Likewise, we find no correlation between the ADF and either the diameter or relative diameters measured here.  

Figures \ref{fig_mosaic_elliptical} and \ref{fig_mosaic_bipolar} present montages of images for 58 planetary nebulae from this sample.  For the objects that do not appear (18 objects), we were unable to find adequate images from which to judge their morphologies, i.e., they are all compact and we cannot distinguish their structure.  The presence of clear morphological structure in Figures \ref{fig_mosaic_elliptical} and \ref{fig_mosaic_bipolar} is striking, especially the preponderance of elliptical, bi- and multipolar morphologies.  

Table \ref{table_properties} presents a list of spectral classifications for the central stars in our sample, when known.  It also indicates the central stars known or presumed to be a binary, or derived from a binary merger.  Given the few objects related to binary central stars, little can be said based upon our sample.  However, \citet{corradietal2015} and \citet{jonesetal2016} have argued recently for a connection between binarity and high ADFs in planetary nebulae.  On the other hand, 32 of the 48 central stars in Table \ref{table_properties} appear to have emission lines, presumably indicating significant winds.  Approximately half of these are of the \lq\lq wels'' type, which has been brought into question recently \citep[e.g.,][]{weidmannetal2015, basurahetal2016}.  We emphasize, however, that the kinematics of the \ion{C}{2} $\lambda$6578 line do not appear to be congruent with the expectations of ionization equilibrium, unlike the conclusions of \citet{basurahetal2016} for the \ion{C}{4} $\lambda$5801 line (based upon its spatial extent).

From the foregoing, both the well-defined morphology and the limited range in H$\alpha$ line widths (Figures \ref{fig_fwhm_o3_c2}, \ref{fig_mosaic_elliptical}, and \ref{fig_mosaic_bipolar}) may be interpreted as indications that the sample of planetary nebulae considered here is composed of young objects.  Our grid of models indicates that high electron density, high stellar luminosity, and the combination of compact size and low stellar temperature all favor high surface brightness in the \ion{C}{2} $\lambda$6578 line.  All of these characteristics are expected to be more common for young planetary nebulae, which may help explain the composition of our sample.   

\subsection{Comparison with previous work}
\label{section_general_considerations}

Previous work has also found that the spatial distribution of the emission from permitted lines is more centrally-concentrated than that from collisionally-excited emission from the same parent ions \citep[cf.][]{barker1982, barker1991, liuetal2000, garnettdinerstein2001, luoliu2003, tsamisetal2008, corradietal2015, jonesetal2016, garciarojasetal2016a, garciarojasetal2016b}.  These were detailed studies of 11 individual objects (A 46, A 63, Hf 2-2, M 1-42, NGC 2392, NGC 6720, NGC 5882, NGC 6153, NGC 6778, NGC 7009, and Ou 5).  Our analysis is unable to expand upon these results.  We note, however, that studies that compare the spatial distributions of permitted and collisionally-excited lines and those that compare the kinematics of these two classes of lines are complementary.  The spatial distributions probe the projections of the volumes of the emitting plasmas in the plane of the sky whereas the kinematics probe the projections of these volumes along the line of sight.  In principle, both methods should come to the same conclusions.  

As regards the discrepant kinematics of the \ion{C}{2} $\lambda$6578 line, similar results have been found on a few occasions previously.  \citet{richeretal2013} found that the kinematics of one component of the emission from permitted lines of \ion{C}{2}, \ion{N}{2}, \ion{O}{2}, and \ion{Ne}{2} in NGC 7009 did not coincide with expectations based upon ionization equilibrium.  \citet{penaetal2016} find similar results for Cn 1-5 and PC 14.  \citet{otsukaetal2010} found that kinematics of the \ion{O}{2} lines in BoBn 1 were not congruent with the ionization structure, but that the kinematics of the \ion{C}{2}, \ion{N}{2} and \ion{Ne}{2} lines did agree with expectations.   \citet{barlowetal2006} also found that the kinematics of the \ion{O}{2} lines in NGC 6153 did not conform to the expectations of ionization equilibrium.  We consider the results for IC 418 inconclusive \citep{sharpeeetal2004}, since those observations could confuse the kinematic and ionization structures in this object.  From a statistical point of view, we find that the kinematics of the \ion{C}{2} $\lambda$6578 line do not agree with the expectations based upon ionization equilibrium in a chemically homogeneous nebula for our sample of 76 planetary nebulae (and 83 lines of sight).  

\begin{deluxetable}{lccllccl}
\tabletypesize{\footnotesize}
\tablewidth{0pt}
\tablecaption{Central star properties}
\tablehead{
\colhead{object$^a$} & \colhead{Sp. type} & \colhead{binary} & \colhead{Ref.} & \colhead{object$^a$} & \colhead{Sp. type} & \colhead{binary} & \colhead{Ref.} \\
}
\startdata
Cn 1-5       & [WO4]pec          &      & 17     &  M 2-14       & wels              &      & 19     \\
H 1-11       & wels              &      & 17     &  M 2-16       & [WO2-3]           &      & 25     \\
H 1-24       & wels              &      & 17     &  M 2-19       &                   & yes  & 24     \\
H 1-33       &                   & yes  & 24     &  M 2-27       & [WC4]:            &      & 25     \\
H 1-56       & wels              &      & 23     &  M 2-29       & Of(H)             &      & 26     \\
Hb 12(a)     & wels?/B[e]?       & yes  & 12,21  &  M 2-30       & wels              &      & 23     \\
He 2-436     & [WC4]             &      & 17     &  M 2-31       & [WC4]             &      & 17     \\
Hf 2-2(b)    &                   & yes  & 13     &  M 2-33       & O5f(H)            &      & 22     \\
Hu 1-1       & A?                & yes  & 3,27   &  M 2-39       & wels              &      & 25     \\
IC 2149(c)   & O4f               &      & 4      &  M 2-8        & [WO3]             &      & 19     \\
IC 418(c)    & Of(H)             &      & 7      &  M 3-17       & [WC11]?           &      & 19     \\
IC 4593(j)   & O7                &      & 8      &  M 3-6        & wels              &      & 17     \\
J 900        & wels              &      & 9      &  Me 2-2       & Of                &      & 6      \\
K 3-66       & cont.             &      & 1      &  NGC 1535(a)  & O(H)              & yes  & 7,14   \\
M 1-14       & OB                &      & 27     &  NGC 6543(a)  & Of-WR(H)          &      & 29     \\
M 1-19       & wels?             &      & 17     &  NGC 6567     & wels              &      & 9      \\
M 1-30       & wels              &      & 17     &  NGC 6629     & [WC4]?            &      & 17     \\
M 1-31       & wels              &      & 9      &  NGC 6778     & cont.             & yes  & 10,26  \\
M 1-35       & wels              &      & 25     &  NGC 6826     & O6fp              & yes$^a$  & 2,5,28 \\
M 1-37       & [WC11]?           &      & 20     &  NGC 7662(c)  & UV emission lines &      & 10     \\
M 1-46       & Of(H)             &      & 18     &  PC 12        & OB                &      & 27     \\
M 1-6        & emission-line     &      & 27     &  Ps 1         & sdO               &      & 16     \\
M 1-61       & wels              &      & 17     &  Sp 4-1       & wels              &      & 9      \\
M 1-65       & [WR]-Of           &      & 11     &  Vy 1-1       & O(H)              &      & 15     \\
\enddata
\label{table_properties}
\tablenotetext{a}{The central star is a binary merger remnant.}
\tablecomments{References:  1--\citet{kohoutek1969}; 2--\citet{smithaller1969}; 3--\citet{kaler1976}; 4--\citet{heap1977}; 5--\citet{lawritter1983}; 6--\citet{allerkeyes1987}; 7--\citet{mendezetal1988}; 8--\citet{bianchidefrancesco1993}; 9--\citet{tylendaetal1993}; 10--\citet{feibelman1994}; 11--\citet{kondrat'eva1994}; 12--\citet{hyungaller1996}; 13--\citet{lutzetal1998}; 14--\citet{ciardulloetal1999}; 15--\citet{napiwotzki1999}; 16--\citet{rauchetal2002}; 17--\citet{ackerneiner2003}; 18--\citet{handler2003}; 19--\citet{gornyetal2004}; 20--\citet{gesickietal2006}; 21--\citet{hsiaetal2006}; 22--\citet{hultzschetal2007}; 23--\citet{gornyetal2009}; 24--\citet{miszalskietal2009}; 25--\citet{depewetal2011}; 26--\citet{miszalskietal2011}; 27--\citet{weidmanngamen2011}; 28--\citet{demarcoetal2015}; 29--\citet{weidmannetal2015}}
\end{deluxetable}

We caution, however, that our result does not imply that the kinematics of the \ion{C}{2} $\lambda$6578 permitted line  never agree with ionization equilibrium.  In NGC 7009, \citet{richeretal2013} deduced that there were two emission components emitting in the \ion{C}{2}, \ion{N}{2}, \ion{O}{2}, and \ion{Ne}{2} permitted lines, of which the kinematics of only one did not conform to the expectations of ionization equilibrium.  We are unable to discern detail so fine with our data.  Since we find that the kinematics of the \ion{C}{2} $\lambda$6578 line commonly does not agree with the expectations based upon ionization equilibrium, it is likely that some important fraction of the plasma emitting the \ion{C}{2} $\lambda$6578 line (rather than all of it) in many of the objects in our sample has kinematics that do not match expectations based upon ionization equilibrium in a chemically-homogeneous nebular plasma.  Given the example of multiple plasma components in NGC 7009, our results would imply that this result is common.  

As argued by various studies dating back to \citet{liuetal2000}, the result that an important fraction of the \ion{C}{2} $\lambda$6578 emission from many planetary nebulae has kinematics or a spatial distribution that are discrepant with respect to the ionization structure favors solutions to the abundance discrepancy problem that do not depend upon a chemically-homogeneous structure for these nebulae.  That is, the abundance discrepancy may arise because there are multiple plasma components.  However, should this be the case, it merely displaces the problem, since it is then necessary to explain how multiple plasma components might arise in many planetary nebulae.  Unless our sample is unfortunately biased, our observations indicate that multiple plasma components are common in planetary nebulae, so it is important to understand how they arise if we hope to fully understand ionized plasmas throughout the universe and to correctly interpret the chemical abundances derived from their observation.  

Several objects are observed to have chemically distinct plasma components, A30, A58, and A78 being the best-known \citep{jacobyford1983, lauetal2011, fangetal2014}, and are thought to be the result of either very late thermal pulses or novae.   Sakurai's object (V4334 Sgr) is another object that has undergone a very late thermal pulse.  It has produced a significant amount of dust since its outburst, but it is unknown if its composition differs from that of the surrounding nebula \citep{chesneauetal2009, hinklejoyce2014}.  Other similar objects include FG Sge, IRAS15154-5258, IRAS18333-2357, and perhaps CK Vul and nova V458 Vul \citep{gillettetal1989, gonzalezetal1998, hajduketal2007, wessonetal2008}.  While these objects clearly present chemically-distinct plasma components, it is unknown whether the means suggested to produce them occur sufficiently frequently to account for their presence \citep{lauetal2011}.  To account for a sample as large as that studied here or the other objects discussed in this section, presumably additional mechanisms are required to produce chemically-distinct plasma components.

\section{Conclusions}

We have studied the kinematics of the \ion{C}{2} $\lambda$6578 line along 83 lines of sight in 76 individual planetary nebulae using data from the SPM Kinematic Catalogue of Planetary Nebulae \citep{lopezetal2012}.  This sample includes all of the objects known to present secure detections of the \ion{C}{2} $\lambda$6578 line among the more than 600 planetary nebulae in the SPM Catalogue.  Since the selection criteria for the SPM catalogue did not include knowledge of any chemical abundances or the presence of the \ion{C}{2} $\lambda$6578 line, our sample was selected blindly for the purposes of this study.  The high resolution spectra in the SPM Catalogue were mostly acquired at the OAN-SPM using the MES attached to the 2.1\,m telescope and allow us to measure line widths for the [\ion{N}{2}] $\lambda\lambda$6548,6583, \ion{He}{2} $\lambda$6560 (when present), H$\alpha$, and \ion{C}{2} $\lambda$6578 emission lines.  From these spectra, we construct one-dimensional spatial profiles of all objects in all lines, from which we measure the spatial extents of each object in each line.  

We find that the velocity width of the \ion{C}{2} $\lambda$6578 line is usually narrower than H$\alpha$, almost always narrower than [\ion{N}{2}] $\lambda$6583, but almost always broader than \ion{He}{2} $\lambda$6560.  For this sample, the line widths of the H$\alpha$, \ion{C}{2} $\lambda$6578, and [\ion{N}{2}] $\lambda$6583 lines all are systematically larger when the \ion{He}{2} $\lambda$6560 line is present, reflecting the same acceleration of the nebular shells as the central star evolves reported by \citet{richeretal2008}.  To account for this evolutionary effect, we consider the line widths relative to the line width of H$\alpha$.  

We find that the kinematics of the \ion{C}{2} $\lambda$6578 line are discrepant.  This conclusion is based upon the results of a grid of models that consider both the recombination and fluorescence components of the \ion{C}{2} $\lambda$6578 line.  These models indicate that \ion{C}{2} $\lambda$6578 emission arises throughout the zone from which H$\alpha$ arises, though its sampling of this volume depends upon the temperature of the central star.  Comparing the line width distributions of the \ion{C}{2} $\lambda$6578 and H$\alpha$ lines, the distribution of relative line widths of the \ion{C}{2} $\lambda$6578 and [\ion{O}{3}] $\lambda$5007 lines, and the median value of the distribution of \ion{C}{2} $\lambda$6578 relative line widths, we repeatedly find that the distribution of \ion{C}{2} $\lambda$6578 line widths or relative line widths is shifted to lower values than expected.  Statistically, the kinematics of the \ion{C}{2} $\lambda$6578 line do not match expectations based upon ionization equilibrium in a chemically-homogeneous plasma.      

We find no correlation between nebular kinematics and the ADF observed for oxygen.  However, we find that the line widths in H$\alpha$ for the objects in our sample are limited, with no large line widths, and that the nebular morphology is complex for those objects for which adequate imaging data exists, both of which could be explained if the objects in which the \ion{C}{2} $\lambda$6578 line is observed are young planetary nebulae.  The trends in the \ion{C}{2} $\lambda$6578 surface brightness from our grid of models is also compatible with this explanation.  The incidence of WR- and wels-type central stars is 66\% among the 48 planetary nebulae in our sample whose central stars have classifications.  About half are of the wels type.  Whether the central stars have a direct influence on the kinematics we find for the \ion{C}{2} $\lambda$6578 line requires further investigation.  

Our results for the kinematics of the emission from the \ion{C}{2} $\lambda$6578 line agree with previous findings regarding the spatial distributions or kinematics of permitted lines \citep{barker1982, barker1991, liuetal2000, garnettdinerstein2001, luoliu2003, barlowetal2006, tsamisetal2008, otsukaetal2010, richeretal2013, corradietal2015, jonesetal2016, garciarojasetal2016a, penaetal2016}.  All of these studies, however, pertain to only 14 objects in total whereas our sample is much larger  (and includes 5 objects in common with these studies).  Considering all data, it appears that it is common that at least an important fraction of the emission of the \ion{C}{2} $\lambda$6578 line in many planetary nebulae arises from a zone more internal than expected based upon ionization equilibrium in a chemically-homogeneous nebular plasma.  Hence, our results argue that multiple plasma components may be common in planetary nebulae.




\acknowledgments

We thank the referee for many helpful comments and suggestions.  MGR thanks W. Henney and G. Ferland for very insightful discussions.  We acknowledge financial support from UNAM-DGAPA grants IN 110011 and IN 108416 and CONACyT grant 178253.  We thank the daytime and night support staff at the OAN-SPM for facilitating and helping obtain our observations, in particular Gabriel Garc\'\i a (deceased), Francisco Guill\'en, Gustavo Melgoza, Salvador Monrroy, and Felipe Montalvo.  We also thank the time allocation committee for awarding time over many years to this ongoing project.  This research has made use of the SIMBAD database, operated at CDS, Strasbourg, France \citep{wengeretal2000}.  Funding for SDSS-III has been provided by the Alfred P. Sloan Foundation, the Participating Institutions, the National Science Foundation, and the U.S. Department of Energy Office of Science. The SDSS-III web site is http://www.sdss3.org/.
SDSS-III is managed by the Astrophysical Research Consortium for the Participating Institutions of the SDSS-III Collaboration including the University of Arizona, the Brazilian Participation Group, Brookhaven National Laboratory, Carnegie Mellon University, University of Florida, the French Participation Group, the German Participation Group, Harvard University, the Instituto de Astrofisica de Canarias, the Michigan State/Notre Dame/JINA Participation Group, Johns Hopkins University, Lawrence Berkeley National Laboratory, Max Planck Institute for Astrophysics, Max Planck Institute for Extraterrestrial Physics, New Mexico State University, New York University, Ohio State University, Pennsylvania State University, University of Portsmouth, Princeton University, the Spanish Participation Group, University of Tokyo, University of Utah, Vanderbilt University, University of Virginia, University of Washington, and Yale University.



Facilities: \facility{OAN-SPM, Hubble Legacy Archive, SDSS, Simbad}.

\end{document}